%% file: main.tex
\documentclass[12pt]{article}

\usepackage{amssymb,amsmath,amsfonts,eurosym,geometry,ulem,graphicx,caption,color,setspace,sectsty,comment,footmisc,caption,natbib,pdflscape,subfigure,array,threeparttable,adjustbox,rotating,authblk,booktabs,dcolumn,tabularx,subcaption,float,titling,enumitem,xcolor,longtable,amsthm,setspace,enumitem,titlesec}
\usepackage{chngcntr}
\usepackage[none]{hyphenat}
\usepackage[section]{placeins}
\usepackage[title]{appendix}

 \usepackage{newpxtext}
 \usepackage{newpxmath} 

 \usepackage[T1]{fontenc}
 \usepackage{microtype}
\usepackage[colorlinks=true, urlcolor=darkblue, citecolor=darkblue, linkcolor=darkblue]{hyperref}
\definecolor{darkblue}{rgb}{0.0, 0.0, 0.75}

\newcolumntype{d}[1]{D{.}{.}{#1}}

\newcolumntype{L}[1]{>{\raggedright\let\newline\\arraybackslash\hspace{0pt}}m{#1}}
\newcolumntype{C}[1]{>{\centering\let\newline\\arraybackslash\hspace{0pt}}m{#1}}
\newcolumntype{R}[1]{>{\raggedleft\let\newline\\arraybackslash\hspace{0pt}}m{#1}}

\begin{document}

\begin{titlepage}
\title{\Large{\textbf{The Dispossessed: Large-Scale Land Acquisitions, \\
Elite Capture, and Dissent in Africa}}}
\author{\large{Jonathan Dries}\thanks{Department of Economics and Finance, LUISS Guido Carli University, Rome, Italy (\textcolor{darkblue}{jvdries@luiss.it}). \\ I am grateful to Luigi Pascali, Mounu Prem, and Valeria Rueda for their guidance and support. I thank Koen Schoors for helpful comments. The title is a nod to Ursula K. Le Guin's 1974 novel. All errors are my own.}}
\date{\large{\textsc{June 2026}}}
\maketitle
\begin{abstract}
\noindent Over the past two decades, millions of hectares of land in Africa have been transferred to investors, raising fears of displacement and conflict. This paper estimates the causal impact of large-scale land acquisitions (LSLAs) on local dissent by comparing successfully implemented projects to a control group of exogenously failed deals. Using staggered difference-in-differences estimators across 1,391 geocoded deals, I find that LSLAs cause a sustained increase in civic unrest of 158\% relative to the pre-treatment mean. Protest responses are strongest among domestic investors acquiring community or state land for food-crop production, pointing to local dispossession and domestic elite capture. Integrating media, survey, and electoral data consistent with this hypothesis, I document parallel shifts in property-rights media discourse, an erosion of traditional authority, and broader electoral mobilization in affected constituencies.
\vspace{3mm}
\vspace{3mm}

\bigskip
\end{abstract}
\setcounter{page}{0}
\thispagestyle{empty}
\end{titlepage}
\pagebreak \newpage
\doublespacing

\input{sections/I._Introduction}

\input{sections/II._Background}

\input{sections/III._Data_and_Measurement}

\input{sections/IV._Causal_Evidence}

\input{sections/V._Mechanisms}

\input{sections/VI._Conclusion}

\clearpage

\singlespacing

\setlength\bibsep{5pt}
\bibliographystyle{apalike}
\bibliography{references}
\doublespacing

\clearpage

\input{sections/VII._Appendix}

\end{document}

%% file: sections/I._Introduction.tex
\section{Introduction}

\noindent In the first two decades of the 21st century, an area of land roughly the size of California was leased or sold to private investors in the developing world. While the economic logic of this global land rush was well documented,\footnote{The surge in global land acquisitions was catalyzed by the 2007--2008 world food price crisis. A confluence of extreme weather events, rising energy costs, biofuel mandates, and protective export bans led to a spike in agricultural commodity prices. This prompted net-food-importing countries and institutional investors to acquire arable land overseas as a hedge against future supply shocks and price volatility \citep{cotula2009grab,byerlee2013growing}.} its political and social consequences remain contested \citep{deschutter2011grabbing}. This paper estimates the longer-run causal impact of these large-scale land acquisitions (LSLAs) on local stability in Africa. Across the continent, the transfer of poorly documented customary land to formal investors has triggered international concern over forced displacement and state-backed expropriation. Although contemporary policy discussions frequently emphasized transnational ``land grabs,'' empirical evidence isolating the medium-term effects of LSLAs and the related mechanisms has remained limited \citep{balestri2021land}.

This paper argues that the economic stakes of these transfers depend on an institutional friction in Africa. Across large parts of the continent, land is governed by a fragmented dual tenure system in which communities hold customary rights over ancestral land while national governments simultaneously assert formal statutory ownership over all unregistered land \citep{peters2004inequality,platteau2009institutional,boone2014property}. Since customary land often lacks cadastral registration, state authorities can lease it to private investors as legally ``empty'' domain, potentially bypassing traditional rights without community consent or fair compensation. Communities lacking formalized institutional defenses are vulnerable to this type of external encroachment and resource extraction \citep{platteau2002gradual,feir2024when}, with theoretical models suggesting that imposing exclusive private property regimes over communal land can fracture social cohesion and reduce collective welfare \citep{platteau1996evolutionary,eswaran2023wrongs}. Therefore, when politically connected domestic elites or transnational corporations use this ambiguity to appropriate customary land by administrative fiat, do affected communities mobilize against it?

A central challenge in evaluating the consequences of LSLAs is the endogeneity of site selection. Investors systematically target high-value, arable land located near strategic infrastructure, water resources, and markets. These same geographic fundamentals independently predict local economic activity, population density, and historical conflict. To overcome this selection bias, I construct an identification strategy exploiting a localized control group of failed or abandoned land deals across 38 African countries between 1997 and 2025. Following the logic of using unimplemented projects as counterfactuals \citep{greenstone2010agglomeration,donaldson2018railroads}, I compare successfully implemented LSLAs from the Land Matrix database against transactions that failed due to exogenous investor-side frictions before any operational activity occurred.\footnote{To verify that the control group is not endogenous to the outcome of interest, I exclude 19.1\% of failed or abandoned candidate controls in the final analytical sample when the failure record contains deal-specific evidence of community resistance.} Moreover, robustness exercises using a not yet treated control group from later finalized deals yield similar results. Combining this with high-resolution conflict data from the Armed Conflict Location \& Event Data Project (ACLED), I employ a staggered difference-in-differences imputation estimator \citep{borusyak2024revisiting} to recover event-study estimates of the average treatment effect on the treated.

The baseline empirical specification documents a sustained increase of approximately 1.48 protest or riot events per location-year in the affected neighborhood, more than doubling the local protest count relative to the pre-treatment mean of 0.94 events per year. To contextualize this magnitude, the effect is roughly four to five times larger than the benchmark impact of a negative rainfall shock on conflict in Africa estimated by \citet{miguel2004economic}. Event-study estimates confirm that parallel pre-trends hold, with unrest beginning to rise only after the formal acquisition. Furthermore, the protest estimates remain robust to substantial violations of the parallel trends assumption when evaluated using the bounds approach of \citet{rambachan2023credible}.

Heterogeneity analyses reveal that this mobilization is not a uniform response to investment. When separating deals by investor origin, the escalation in unrest is most pronounced and dynamic among deals led by domestic actors. Transnational investors also generate positive protest responses, but their effects are smaller and less clearly sustained. The same separation applies to crop type: food-crop acquisitions, which directly displace subsistence livelihoods, generate the bulk of the response, while biofuel and other non-food commercial crops do not. This double divergence aligns with the institutional constraints these actors face: domestic elites can leverage local political capital to bypass community consultation mechanisms, while international agribusinesses are increasingly bound by global ESG standards and reputational scrutiny.

A former land-ownership split, identified separately under a later-treated comparison design as prior-owner information is sparser among failed controls, complements this picture. Acquisitions of community or indigenous-titled land generate large and persistent protest increases, and acquisitions of formerly state-administered land also produce a robust protest response. Together with the domestic, food-crop concentration above, these patterns point toward a specific configuration as the locus of unrest: the commercialization of food crops on customary or state-administered land by politically connected domestic actors, rather than a generic response to investment.

These patterns support a domestic elite-capture interpretation as opposed to the prevailing transnational ``land grab'' narrative. The evidence validates theoretical models of African state-building in which politically connected domestic actors exploit the institutional ambiguity of dual-tenure systems. By leveraging state facilitation to execute top-down formalizations, these elites can appropriate customary land without community consent, which appears to trigger sustained local resistance.

Three sources of complementary evidence speak to the mechanisms underlying this mobilization. First, an EA-round panel built from six rounds of Afrobarometer microdata reveals an erosion of traditional authority: exposure to LSLAs reduces trust in traditional leaders and lowers contact with traditional leaders as community problem-solvers, indicating that communities specifically blame the local actors who either facilitated the transfer or failed in their protective function. This custodial-failure channel is consistent with an erosion of the social buffer through which land grievances might otherwise be contained, leaving communities with fewer alternatives to redirect grievances away from public protest.

Second, I examine media coverage using the GDELT Global Knowledge Graph, which shows a concentrated escalation in property-rights, corruption, and agricultural coverage near LSLA sites, pointing to grievance-specific media attention. Property-rights, corruption, and agriculture-themed coverage all rise near implemented deals relative to the counterfactual, illustrating how localized land disputes capture media attention and seem to escalate into broader public narratives.

Third, using constituency-level electoral data from the Comparative Legislative Elections Archive (CLEA), I document a strong electoral response: opposition vote shares and voter turnout rise in constituencies near implemented deals, while incumbent punishment is weaker and less stable. Operating downstream on a slower four-to-five-year electoral cycle, these results show how affected communities translate their institutional frustrations into stronger opposition support and higher turnout. Together, these three channels suggest a political economy in which dispossession and the weakening of customary authority translate into collective action through the redirection of grievances into the media and formal political arena.

These findings carry specific implications for the political economy of agricultural development. They challenge the evolutionary theory of land rights, which posits that rising scarcity drives an organic, welfare-enhancing transition from customary to formal tenure \citep[see][]{platteau1996evolutionary}. Instead, this evidence suggests that when formalization is imposed top-down by domestic elites it may bypass community consent, thereby generating sustained unrest. International frameworks that focus overwhelmingly on imposing ESG-style due diligence on transnational investors may leave a domestic regulatory gap, enabling elite capture that drives localized mobilization.

\paragraph{Related Literature.} This paper contributes to four interconnected strands of literature. First, the paper offers new empirical evidence on the socio-economic impacts of the global land rush \citep{cotula2009grab,deschutter2011grabbing,nolte2014zambia,arezki2015drives,kleemann2015welfare,molebatsi2019land,stojanovic2020large}. Previous literature has noted a spatial correlation between large-scale land acquisitions and organized, lethal violence \citep{balestri2021land}. Evaluating the local externalities of large-scale agricultural investments poses known empirical challenges, particularly regarding endogenous site selection and heterogeneous treatment effects \citep{goodman2021difference}. By leveraging a large sample with a localized control group composed of failed implementations \citep{greenstone2010agglomeration,donaldson2018railroads} and adopting a staggered imputation estimator \citep{borusyak2024revisiting}, I attempt to isolate the causal effect of land transfers from their endogenous placement. I then move beyond lethal violence to examine the broader political economy of LSLAs, integrating survey and electoral data to show how these deals erode traditional authority and mobilize electoral backlash.

Second, I contribute to the literature on property rights, tenure security, and agricultural investment \citep{besley1995property,platteau1996evolutionary,jacoby2002expropriation,deininger2008overlapping,goldstein2008profits,fenske2011tenure,fenske2014trees,huntington2021does}. Building on the institutional background described above, I provide causal evidence that top-down formalization by domestic elites triggers sustained civic mobilization. This adds to ongoing debates regarding whether policy-induced formalization improves or worsens rural welfare \citep{atwood1990registration,deininger2001institutions,bromley2009formalising,goldstein2018formalization}. Specifically, my findings provide modern empirical support for theoretical concerns that imposing individualized private property regimes over customary, communal land can fracture social cohesion and generate localized unrest \citep{eswaran2023wrongs}.

Third, the paper speaks to the political economy literature on resource shocks and conflict \citep{dell2010mita,dube2013commodity,berman2017mine,christensen2019concession,crost2020extractive,sandi2025mining,rigterink2025mining}. A substantial body of evidence documents associations between global resource demand, local wealth, and the onset of armed violence, often intertwined with historical borders and fiscal capacity \citep{eifert2010political,cogneau2014borders,cogneau2021fiscal}. By building on advances in the disaggregated analysis of localized shocks \citep{harari2018conflict}, I show that institutional capture, specifically the strategic exploitation of tenure ambiguity to appropriate customary land, is associated most clearly with civic mobilization in my setting.

Finally, I bridge the historical literature on legal plurality with the political economy of traditional authority and accountability in Sub-Saharan Africa \citep{chanock1991peculiar,berry1993permanent,blattman2010civil,baldwin2015chiefs}. Within dual-tenure environments, institutional ambiguity creates structural vulnerabilities \citep{peters2004inequality,cotula2011land,hall2011land}. I provide micro-level evidence for theoretical mechanisms regarding how empowered local actors and chiefs exploit this ambiguity to capture rents at the expense of local development \citep{acemoglu2014chiefs,boone2014property,boone2019legal}. I find that this erosion of customary authority reduces trust in traditional leaders and coincides with higher opposition support and turnout, linking asset dispossession to electoral mobilization in weak-institution settings \citep{ferraz2008exposing,brollo2013political,burgess2015democracy,bobonis2016vulnerability}. In this context, the findings offer a contemporary African parallel to the historical dispossession of Indigenous nations in North America, illustrating how the attenuation of customary property rights and the overriding of local institutional capacity can serve as catalysts for dispossession and long-term marginalization \citep{farrell2021effects,carlos2022indigenous,feir2024when}.

The remainder of the paper proceeds as follows. Section II provides the institutional background and derives empirical predictions. Section III describes the data, empirical strategy, and estimation framework. Section IV presents the causal evidence, robustness checks, and heterogeneity analyses. Section V examines the mechanisms using media, survey, and electoral data. Section VI concludes.

%% file: sections/II._Background.tex
\section{Institutional Background}
\label{sec:background}

\subsection{The Global Land Rush}

The phenomenon of Large-Scale Land Acquisitions (LSLAs) accelerated following the global food and financial crises of 2007--2008. Driven by rising agricultural commodity prices, new biofuel mandates in the Global North, and capital seeking durable, inflation-proof assets, a diverse array of foreign and domestic actors sought to acquire large tracts of arable land \citep{dejanvry2008global,deininger2011rising}. Sub-Saharan Africa emerged as the primary target for this global land rush. The continent was attractive due to the perception of abundant, ``idle'' land, favorable agro-ecological endowments, and relatively weak or pliable formal land governance frameworks \citep{cotula2009grab}. While early political narratives characterized these investments as a mutually beneficial transfer of technology and capital to capital-starved agrarian regions \citep{worldbank2011rising}, critical observers highlighted the asymmetries in bargaining power between international agribusiness and local smallholders \citep{deschutter2011grabbing}.

\begin{figure}[t]
    \centering
    \includegraphics[width=0.85\textwidth]{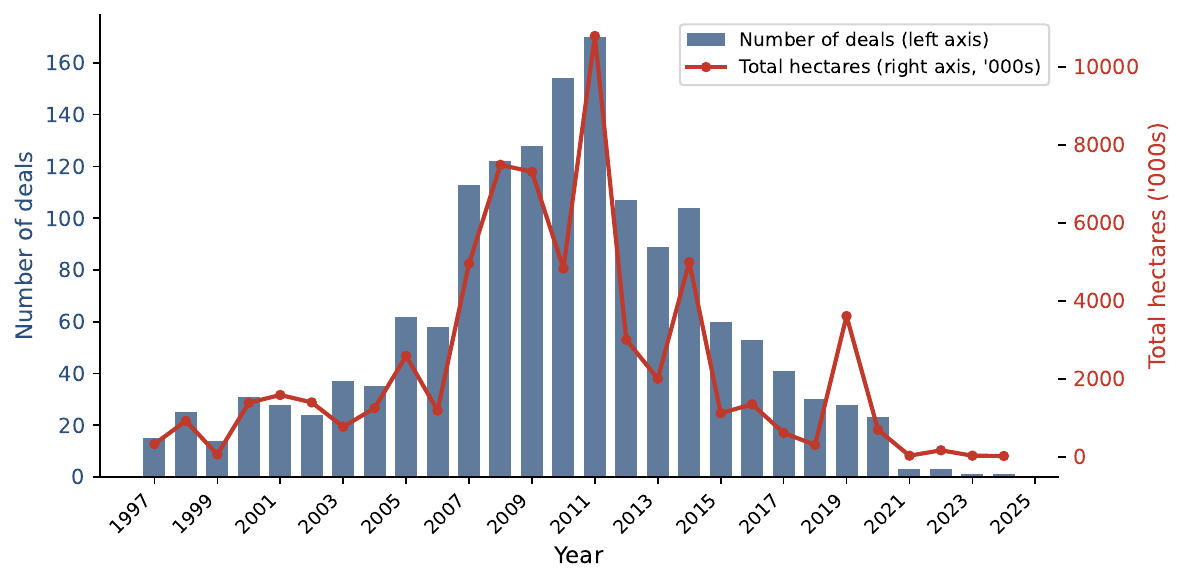}
    \caption{Large-Scale Land Acquisitions in Africa, 1997--2025}
    \caption*{\textit{Notes.} Annual number of newly concluded land deals (bars, left axis) and total hectares transacted in thousands (line, right axis) across the 38-country analytical sample, 1997--2025. Source: Land Matrix database, author's calculation.}
    \label{fig:lsla_trends}
\end{figure}

Figure \ref{fig:lsla_trends} illustrates the evolution of LSLA activity across Africa between 1997 and 2025, measured by the annual number of newly concluded deals and total hectares transacted. The figure documents the acceleration peaking around 2007--2008, coinciding with the commodity price crisis, followed by a gradual decline. This pattern is broadly consistent with prior qualitative accounts \citep{cotula2009grab} and indicates that the land rush was largely a response to global price signals and not a gradual secular trend.

\subsection{Land Tenure}
\label{sec:landtenure}

A prerequisite for understanding the conflict dynamics surrounding LSLAs is the dual nature of land tenure across much of Sub-Saharan Africa. This institutional structure is rooted in the region's colonial and post-colonial history of state-building \citep{chanock1985law,berry1993permanent}. Historically, colonial administrations pursued a strategy of formally demarcating permanent, exclusive property rights for European settlers and commercial plantations, while relegating the indigenous majority population to flexible, non-exclusive customary tenure systems administered by government-appointed local chiefs \citep{mamdani1996citizen,boone2014property}.

This dichotomy persisted through independence, cementing a fragmented landscape of legal plurality. As \citet{boone2014property} theorizes, post-colonial African states have maintained this dual-tenure architecture as an instrument of political control. By keeping rural land unregistered and legally ambiguous, the central state limits the emergence of an independent, propertied rural middle class while retaining the ability to allocate land as political patronage. While local communities held ancestral and customary rights to the land they farmed, fallowed, and grazed \citep{sjaastad1997indigenous}, national governments frequently asserted formal, statutory ownership over all ``unregistered'' land \citep{toulmin2000evolving}. Since customary land often relies on oral histories and adaptive, fluid boundaries, it is legally legible to the state only as ``empty,'' ``idle,'' or ``public'' domain \citep{scott1998seeing,peters2004inequality}. State authorities therefore wield the unilateral legal power to lease this theoretically unowned land to investors at scale, bypassing or extinguishing traditional rights without democratic consent or fair compensation.\footnote{This institutional vulnerability has generated a long record of state-sanctioned dispossession. For instance, during the 1970s and 1980s in Tanzania, the state-owned National Agricultural and Food Corporation forcibly displaced tens of thousands of Barabaig pastoralists from the Basotu Plains to establish the Hanang wheat complex \citep{lane1990barabaig}. Because the Barabaig practiced rotational grazing, the government legally classified the land as ``unoccupied,'' extinguishing their customary access overnight. More contemporaneous to the phenomenon of LSLAs, the Ethiopian government leased large tracts of the Gambella region to agricultural conglomerates by redefining ancestral Anuak and Nuer commons as ``terra nullius'' under federal statutory law during the 2000s.}

Within this framework, traditional leaders occupy an institutionally ambiguous position. Customary tenure is administered through local chiefs who are recognized as the legitimate custodians of communal land, but post-colonial states simultaneously integrated these leaders into local administrative hierarchies as low-level functionaries with limited formal authority \citep{mamdani1996citizen}. This dual role places traditional leaders in a structurally constrained position when a large-scale land transfer occurs: they lack the statutory power to overturn state-issued leases, yet communities may hold them responsible for failing to prevent the transfer. This institutional constraint motivates the mechanism analysis in Section \ref{sec:traditional}.

\subsection{Domestic Elite Capture}

While the majority of early critical literature focused on transnational acquisitions by heavily capitalized foreign entities \citep{deschutter2011grabbing}, the empirical reality may be more complex. A substantial proportion of recorded LSLAs are driven by domestic investors \citep{cotula2011land}. These actors (often politically connected elites, retiring military officers, and wealthy urban businesspeople) possess the entrenched patronage networks and localized political capital necessary to navigate community and state land bureaucracies \citep{balestri2021land}.

Three considerations lead me to expect domestic investors to generate more localized unrest than foreign investors. First, domestic elites enjoy an informational advantage regarding the limits and vulnerabilities of local dual-tenure frameworks. Unlike foreign investors who must rely on costly federal assurances and formal legal proxies, domestic actors understand which district commissioners to approach and which customary chiefs may be amenable to negotiation. Second, international agribusinesses are increasingly subject to ESG standards and scrutiny from non-governmental organizations and multilateral lenders, creating reputational constraints on acquiring actively contested community land. Domestic elites operating in domestic capital markets face fewer such external constraints. Third, as \citet{boone2019legal} notes, formalizing land is a political act that alters local power distributions. Domestic investors are better positioned to directly use state enforcement mechanisms to secure newly formalized leases over what was previously customary land.
It may be precisely these customary, community-owned tracts of land that generate the fiercest civic mobilization when their traditional owners are dispossessed.

A different characteristic that might also be relevant to understand community-level impacts is crop composition. Domestic investors acquiring community land tend to cultivate food crops for domestic or regional markets, whereas foreign agribusinesses more commonly focus on plantation crops or biofuels \citep{cotula2009grab}. Land previously used for subsistence agriculture or communal grazing, when converted to commercial food crop production, removes the direct subsistence base of local households, creating material livelihood pressures which may fuel additional grievances.

%% file: sections/III._Data_and_Measurement.tex
\section{Data and Empirical Strategy}
\label{sec:data}

To estimate the causal effect of large-scale land acquisitions (LSLAs) on local conflict dynamics, I integrate georeferenced data on agricultural investments with event-level conflict data across Africa. I spatially match both datasets to construct a deal-level panel suitable for event-study estimation.

\subsection{Land Acquisition Data}

The primary treatment data are sourced from the Land Matrix database, an independent global land monitoring initiative. The Land Matrix Initiative was established in 2011 as a multi-stakeholder partnership between civil society organizations and academic research institutes to create a publicly accessible, continuously updated global registry of large-scale land transactions \citep{landmatrix2026observatory}.

Each deal record is triangulated across multiple source types. The raw database contains over 48,000 source citations globally, spanning media reports, company filings, government gazette announcements, contracts obtained through freedom-of-information requests, and on-the-ground research conducted by partner organizations including GRAIN, Mongabay, and CIRAD. The global database covers 7,671 deals across 78 countries. Of these, 2,349 are targeted at African countries. Among these African deals, approximately 30 million hectares are currently under contract, of which roughly 7 million are in active production.

Beyond location and status, the Land Matrix records a rich set of deal characteristics: intended deal size; legal nature (lease, concession, exploitation permit, outright purchase, or contract farming); prior land ownership regime (state, community, or private); intended production type (food crops, biofuels, livestock, forestry, mining, or renewable energy); investor nationality and classification; and social variables including community consultation status, community reaction, presence of documented land conflicts, and displacement data. This structure allows me to characterize treatment and control deals along observable dimensions and to conduct a range of heterogeneity analyses.

The Land Matrix records five spatial-accuracy labels: \textit{Country}, \textit{Administrative region}, \textit{Approximate location}, \textit{Exact location}, and \textit{Coordinates}. I restrict the sample to the four sub-national labels (i.e., excluding \textit{Country}): \textit{Administrative region}, \textit{Exact location}, \textit{Approximate location}, and \textit{Coordinates}. A robustness check further excluding \textit{Administrative region} yields similar results.

\subsection{Conflict Outcomes}

\paragraph{ACLED.}

Data on conflict and civic unrest are drawn from the Armed Conflict Location \& Event Data Project (ACLED). ACLED provides georeferenced, event-level records of political violence and protest activity across Africa. The panel spans 1997 to 2025, providing several years of pre-treatment data for the majority of deals in the sample. The main text focuses on two ACLED outcomes. First, I measure civic unrest as the combined count of ``Protests'' and ``Riots'' events. Second, I measure all-conflict events as the total count across ACLED event types. A supplementary appendix analysis then decomposes the broader security response into an armed-conflict composite that combines Battles, Explosions/Remote violence, and Violence against civilians, together with the corresponding component outcomes, strategic developments, and fatalities.

To construct the outcome variables, I aggregate conflict events within spatial buffers around each land deal. The primary specification uses a $50\,\text{km}$ buffer, chosen to correspond to the average community and traditional-authority scale discussed in Section~\ref{sec:landtenure}: it captures the catchment of a typical chiefdom or cluster of villages whose customary institutions are most directly affected by the transfer, and it coincides with the buffer used across the paper's outcome families. I use this single unified radius, instead of tailoring the buffer to each outcome, to avoid any appearance of specification search. For robustness, I report a comprehensive buffer sweep at $\{25, 50, 75, 100\}\,\text{km}$ in Appendix Section~\ref{sec:appendix_buffer_sweep}. 

The conflict count $y_{d,t}^r$ for deal $d$ in year $t$ within radius $r$ is defined as:
\begin{equation}
    y_{d,t}^r = \sum_{c \in C_t} \mathbf{1}\{ \text{dist}(c, \text{location}_d) \leq r \}, \quad r = 50\,\text{km}
\end{equation}
where $\mathbf{1}(\cdot)$ indicates whether a conflict event $c$ at time $t$ falls within distance $r$ of the deal's coordinates.

While ACLED's Africa panel nominally begins in 1997, coverage is sparse before 2002. The project's early years rely predominantly on media reports from a narrower set of sources, so the absence of recorded events in the late 1990s is more likely to reflect non-reporting than genuine calm. Because some treated units are assigned pre-treatment windows that overlap with this low-coverage period, the pre-treatment mean could be understated for early cohorts.

Two features of the empirical design mitigate this concern. Country $\times$ year fixed effects absorb any common within-country expansion in reporting intensity, and deal-level fixed effects absorb any time-invariant location-specific reporting bias. Identification therefore comes from within-deal variation around the treatment date relative to country-year averages, as opposed to from cross-cohort comparisons of pre-treatment levels. As an additional empirical check, Appendix Section~\ref{sec:appendix_early_cohort_exclusion} re-estimates the preferred specifications after dropping all deals with a negotiation start year before 2003, restricting the sample to cohorts whose pre-treatment windows fall entirely within the period of mature ACLED coverage. The restricted ATTs remain positive and statistically significant across both outcomes and all three FE/control variants, with magnitudes close to the full-sample baseline.

\paragraph{GDELT.}

The Global Database of Events, Language, and Tone (GDELT) is a massive, open-source initiative that monitors global news media. It spans print, broadcast, and web formats in over 65 languages in near real-time and has become a standard tool for tracking the high-frequency spatial and temporal distribution of civic mobilization, state coercion, and public discourse. To validate the grievance narrative through an independent source, I specifically draw on the GDELT Global Knowledge Graph (GKG) \citep{leetaru2013gdelt}. The GKG processes newswire and online media globally, tagging each article with a set of thematic descriptors from a standardized ontology. I use the GKG v2 geocoded article corpus, which assigns articles to geographic coordinates based on location mentions in the text.

I construct annual counts of articles tagged with three themes of interest (property rights, corruption, and agriculture) within spatial buffers around each LSLA deal, for the period 2015--2025 (the span of reliable GKG geocoded coverage). The outcome variable is the inverse hyperbolic sine of the annual article count. These data are used in Section~\ref{sec:mechanisms} as a narrative validation of the grievance channels.

\subsection{Survey Outcomes}

To examine the mechanism underlying local mobilization, I integrate georeferenced, individual-level survey data from the Afrobarometer. I pool data from Waves 4 through 9, spanning roughly 2008 to 2023, which provides a cross-section of public opinion across the geographic scope of the Land Matrix data.

The Afrobarometer utilizes clustered national probability samples with coordinates at the enumeration area (EA) level. Using the same spatial matching procedure as the conflict data, I construct a $50\,\text{km}$ buffer around each geo-referenced land deal and intersect it with the EA coordinates. Survey respondents whose EA falls within a $50\,\text{km}$ buffer of an LSLA are included in the sample and assigned the status (implemented or failed) of the corresponding deal. For the difference-in-differences analysis in Section~\ref{sec:mechanisms}, the raw individual-level responses are collapsed to an EA $\times$ round panel, yielding 2{,}240 treated EAs (within $50\,\text{km}$ of an implemented deal) and 540 control EAs (within $50\,\text{km}$ of a failed deal only).

\subsection{Electoral Outcomes}

To examine whether LSLAs generate measurable electoral responses, I use constituency-level data from the Comparative Legislative Elections Archive (CLEA), which records vote shares and turnout for national legislative elections across African countries with geocoded constituency boundaries \citep{kollman2019georeferenced,kollman2024clea}. I match each constituency to the nearest LSLA deal within a $50\,\text{km}$ buffer, applying the same spatial scale and treatment-control classification used for the conflict, survey, and media outcomes. The constituency panel spans election years available in CLEA across the sample countries. Robustness to alternative buffer radii is reported in Appendix Section~\ref{sec:appendix_buffer_sweep}.

Table~\ref{tab:mechanisms_data_summary} presents summary statistics for all three mechanism datasets (GDELT GKG, Afrobarometer, and CLEA) side by side, splitting each sample between neighborhoods of implemented deals and neighborhoods of failed deals. The three panels differ in unit of observation and panel structure: GDELT GKG is a deal $\times$ year panel over 2015--2025; Afrobarometer begins as respondent-level survey microdata and is aggregated to an EA $\times$ round panel for estimation; and CLEA is a constituency $\times$ election panel across legislative elections in the sample countries. Across all three sources, the two groups are broadly comparable in demographics and pre-treatment outcome levels, supporting the validity of the failed-deal control group for each mechanism analysis.

\input{tables/01_mechanisms_data_summary}

\subsection{Empirical Strategy}

The empirical challenge in evaluating LSLAs is endogenous site selection \citep{deininger2011rising}. Investors target regions with high agricultural potential, water, and infrastructure access, which are likely correlated with baseline conflict. To address this, I restrict the sample to locations that were formally targeted by investors, following the approach of \citet{greenstone2010agglomeration} and \citet{donaldson2018railroads}.

\paragraph{Sample construction.}

\begin{figure}[htp]
    \centering
    \includegraphics[width=0.95\textwidth]{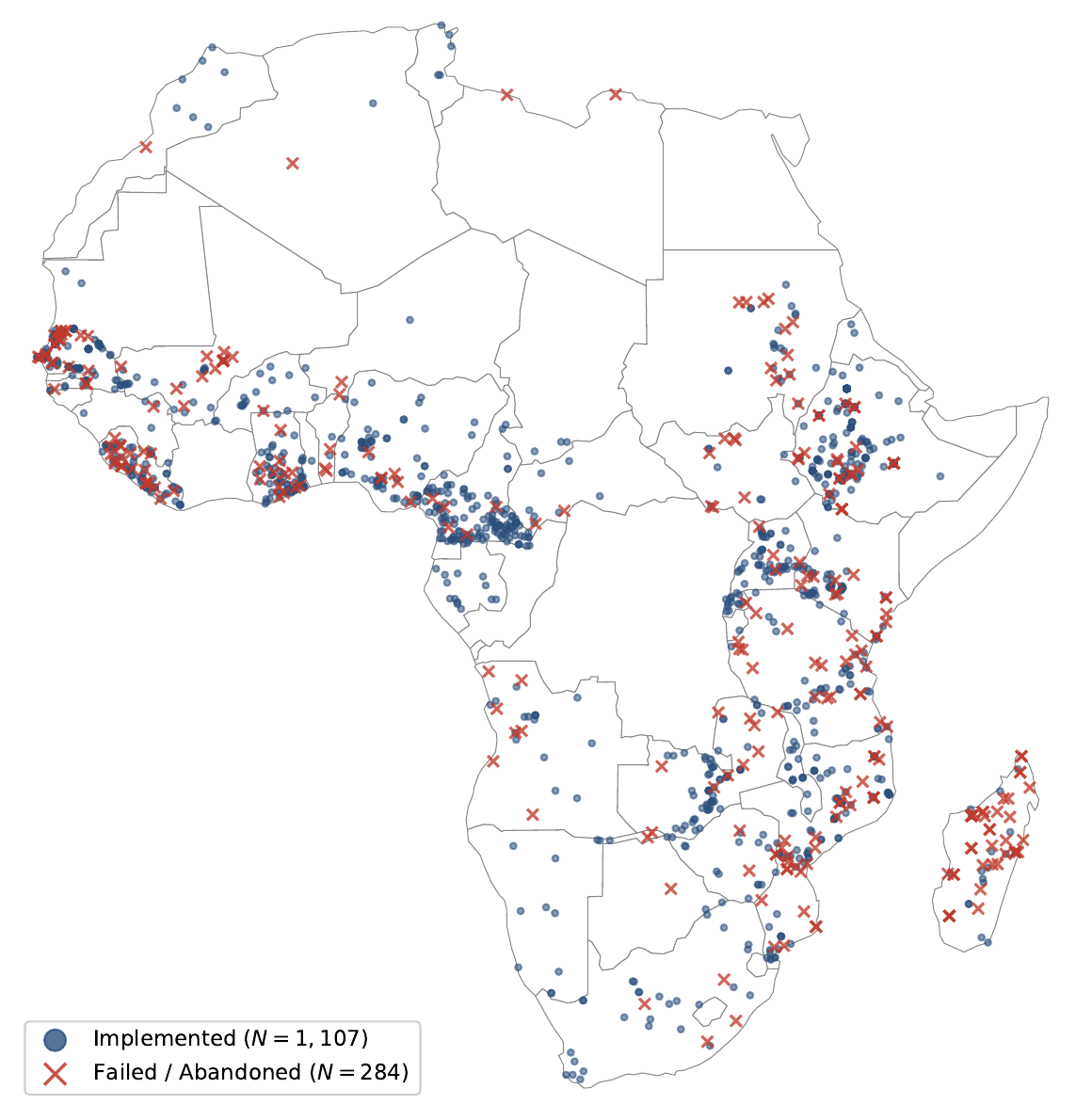}
    \caption{Map of Implemented and Failed LSLA Deals across Africa}
    \caption*{\textit{Notes.} Geographic distribution of implemented deals (blue dots, $N=1,107$) and failed/abandoned baseline-control deals (red crosses, $N=284$) across Africa. Calculations based on Land Matrix data.}
    \label{fig:lsla_map}
\end{figure}

The final analytical sample comprises 1,107 implemented deals (treatment group) and 284 failed or abandoned deals (control group), distributed across 38 African countries. First, deals must carry one of the retained spatial-accuracy labels and valid coordinates. Second, deals must have a recorded negotiation start year to determine treatment timing. In robustness checks, I also use the recorded implementation onset year as an alternative timing for treatment assignment. Third, deals must be unambiguously classified: the treatment group consists of deals that reached contract signature and are ``in operation,'' while the control group consists of deals whose status is ``Failed'' or ``Project abandoned'' before production commenced and whose failure record contains no deal-specific evidence of community resistance. Together, these restrictions yield a sample of 1,391 deals, mapped in Figure \ref{fig:lsla_map}. Table \ref{tab:summary_stats} presents summary statistics for the baseline characteristics of implemented versus failed deals.

\input{tables/02_summary_stats}

\paragraph{Failed deals.}

The use of failed deals as a control group rests on the assumption that successful and abandoned investments share similar unobserved baseline characteristics (e.g., land quality, accessibility). Both groups were deemed viable for investment, but the control group was never operationalized. This allows me to separate the effect of actual land conversion from the mere expectation of investment. The primary identifying assumption is parallel trends: without the completed deals, implemented and failed locations would have followed similar conflict trajectories.

A potential threat to this strategy is endogeneity in deal failure: if deals fail precisely because of pre-existing community resistance, the control group would be selected based on high initial unrest. To mitigate this, I classify failure reasons using three parallel Land Matrix fields: (i) free-text comments on deal status, (ii) the structured \textit{Presence of land conflicts} field, and (iii) the structured \textit{Community reaction} field. I then conduct a supplementary review of unresolved failed deals, checking Land Matrix deal pages, attached sources, and broader web evidence. Following a hierarchical rule that prioritizes community resistance as the most conservative label, I exclude failures where local resistance was a factor. In the final analytical sample, this review removes 67 of 351 failed or abandoned candidate controls, including nine additional resistance-linked failures uncovered by the supplementary source review, leaving a baseline control group of $N=284$ failed or abandoned deals with no deal-specific evidence that community resistance caused the failure. Appendix Table~\ref{tab:control_sample_waterfall} documents the sample construction and failed-control screening procedure.\footnote{I wish to be precise about the content of the ``administrative disputes'' category, which could in principle reflect local institutional resistance in place of exogenous regulatory friction. In this classification, administrative failures refer exclusively to federal- or national-level regulatory actions: license denials or non-renewals by national land ministries or investment promotion agencies, blocks imposed by parliamentary review or presidential decree, revocations triggered by changes in foreign investment law, and complications arising from cross-border contract enforcement in the case of internationally structured deals. These are macro-level shocks that originate at the central-state level and are orthogonal to community-level unrest at the deal site. I explicitly exclude from this category any cases where administrative language co-occurs with evidence of local resistance, assigning such deals the community-resistance label under the hierarchical rule.} Any residual misclassification of endogenously failed deals attenuates the estimates. Deals that fail because of community resistance carry elevated pre- and around-treatment conflict; their inclusion in the untreated pool raises the imputed counterfactual and biases the estimated ATT toward zero. The headline effects should therefore be read as a lower bound with respect to this concern.

\paragraph{Balance and composition.}

Table \ref{tab:summary_stats} reveals two notable observable differences between implemented and failed deals. The control group contains a substantially higher share of transnational investors (84.2\% vs.\ 61.0\%), and failed deals are substantially larger, with a median intended size of 11,950\,ha compared to 3,600\,ha for implemented deals. These patterns are consistent with a selection process in which foreign-led, large-scale projects collapse more often because of higher coordination frictions and capital requirements, not because of local conditions that would threaten identification. Crucially for identification, Country $\times$ year fixed effects absorb country-specific composition shifts in both investor origin and deal size, so the residual identifying variation comes from within-country, within-year comparisons of implemented and failed deals.

Neither compositional difference mechanically biases the event-study estimates. The imputation estimator identifies the ATT from within-deal variation in outcomes over time, conditioning on deal fixed effects that absorb any time-invariant differences in conflict levels. The parallel-trends assumption requires only that the trajectory of conflict in implemented and failed locations would have been similar absent the deal, not that their levels were identical. Panel~D of Table~\ref{tab:summary_stats} reports raw pre-treatment levels on conflict activity, geographic remoteness, and local settlement density. Treated and failed locations differ in raw conflict levels, but Appendix Table~\ref{tab:failure_prediction} shows that, once country fixed effects absorb cross-country differences in Land Matrix documentation intensity, pre-treatment conflict does not significantly predict failure status. I additionally show that the largest protest effects are concentrated in the domestic subset (Section~\ref{sec:heterogeneity_results}), which is more evenly represented across treatment and control.

\subsection{Estimation Framework}
\label{sec:estimation_framework}

To estimate the dynamic impact of large-scale land acquisitions on local conflict, I use the staggered timing of deal implementation across 1,107 target locations compared to 284 counterfactual locations. The primary challenge is the potential for biased estimates in two-way fixed effects (TWFE) models under staggered treatment timing and heterogeneous effects \citep{dechaisemartin2020two,goodman2021difference}. I therefore estimate the impact of LSLAs using a heterogeneity-robust imputation framework.

\paragraph{Imputation estimator.}
I use the imputation estimator of \citet{borusyak2024revisiting}. This approach addresses the negative weighting problem in staggered DiD settings by estimating potential untreated outcomes using only untreated observations: the clean control group of failed deals together with pre-treatment observations of eventually implemented deals. Specifically, I estimate
\begin{equation}
    Y_{it}(0) = \alpha_i + \gamma_{ct} + \varepsilon_{it}
\end{equation}
where $\alpha_i$ are deal fixed effects and $\gamma_{ct}$ are country-year fixed effects, using only untreated observations. I then impute the counterfactual $\hat{Y}_{it}(0)$ for treated units and compute the unit-level treatment effect as
\begin{equation}
    \hat{\tau}_{it} = Y_{it} - \hat{Y}_{it}(0).
\end{equation}
Aggregating over event time $k \geq 0$ recovers positively weighted estimates of the average treatment effect on the treated (ATT). This estimator is robust to arbitrary effect heterogeneity and avoids contamination from previously treated units. The identifying assumption is parallel trends: in the absence of treatment, conflict trajectories in locations with implemented deals would have evolved in parallel to those in the control group,
\begin{equation}
    E[Y_{it}(0) - Y_{i,t-1}(0) \mid D_i = 1] = E[Y_{it}(0) - Y_{i,t-1}(0) \mid D_i = 0].
\end{equation}
The pre-treatment coefficients $\beta_k$ for $k < 0$ provide an empirical test of this assumption.

As a transparency check, Appendix Table~\ref{tab:three_panel_results} reports an additional estimator comparison that places the preferred imputation design alongside alternative modern DiD estimators.

%% file: tables/01_mechanisms_data_summary.tex
\begin{table}[tbp]
\centering
\caption{Mechanism Data Sources and Summary Statistics by Treatment Status}
\label{tab:mechanisms_data_summary}
\begin{threeparttable}
\setlength{\tabcolsep}{8pt}
\begin{tabular}{lccrccr}
\toprule
\textit{Deal status:} & \multicolumn{3}{c}{Implemented} & \multicolumn{3}{c}{Failed} \\
\cmidrule(lr){2-4}\cmidrule(lr){5-7}
 & Mean & SD & $N$ & Mean & SD & $N$ \\
\midrule
\multicolumn{7}{l}{\textbf{Panel A: Afrobarometer}\textit{ (EA $\times$ round, $50\,\text{km}$)}} \\
  Age (years) & 36.66 & (6.34) & 11,024 & 36.68 & (6.02) & 2,789 \\
  Urban (=1) & 0.468 & (0.395) & 11,024 & 0.412 & (0.381) & 2,789 \\
  Education (0--3) & 1.534 & (0.593) & 7,842 & 1.365 & (0.607) & 1,956 \\
  Lived poverty index (0--4) & 1.254 & (0.562) & 5,902 & 1.305 & (0.494) & 1,473 \\
  Trust trad.\ leaders (0--3) & 1.735 & (0.600) & 8,026 & 1.913 & (0.569) & 1,977 \\
  Trust courts (0--3) & 1.583 & (0.547) & 10,006 & 1.652 & (0.534) & 2,478 \\
  Contact trad.\ leader (0--4) & 0.706 & (0.573) & 8,102 & 0.763 & (0.567) & 2,005 \\
\addlinespace[2pt]
  EA-round observations & \multicolumn{3}{c}{11,024} & \multicolumn{3}{c}{2,789} \\
\midrule
\multicolumn{7}{l}{\textbf{Panel B: GDELT GKG} \textit{(deal $\times$ year, $50\,\text{km}$, 2015--2025)}} \\
  $\ln(1+\text{total articles})$ & 4.465 & (2.728) & 10,439 & 4.122 & (2.532) & 2,486 \\
  $\ln(1+\text{property-rights})$ & 1.058 & (1.497) & 10,439 & 0.875 & (1.344) & 2,486 \\
  $\ln(1+\text{corruption})$ & 2.680 & (2.407) & 10,439 & 2.353 & (2.198) & 2,486 \\
  $\ln(1+\text{agriculture})$ & 2.907 & (2.279) & 10,439 & 2.638 & (2.083) & 2,486 \\
\addlinespace[2pt]
  Deal-years & \multicolumn{3}{c}{10,439} & \multicolumn{3}{c}{2,486} \\
  Unique deals & \multicolumn{3}{c}{949} & \multicolumn{3}{c}{226} \\
\midrule
\multicolumn{7}{l}{\textbf{Panel C: CLEA}\textit{ (constituency $\times$ election, $50\,\text{km}$)}} \\
  Opposition vote share & 0.664 & (0.306) & 2,303 & 0.506 & (0.342) & 196 \\
  Incumbent vote share & 0.427 & (0.222) & 1,526 & 0.584 & (0.242) & 141 \\
  Voter turnout & 0.689 & (0.118) & 1,124 & 0.537 & (0.145) & 45 \\
\addlinespace[2pt]
  Constituency-elections & \multicolumn{3}{c}{2,338} & \multicolumn{3}{c}{205} \\
  Unique constituencies & \multicolumn{3}{c}{1,536} & \multicolumn{3}{c}{146} \\
\bottomrule
\end{tabular}
\begin{tablenotes}[para,flushleft]
\footnotesize
\item \textit{Notes.} Summary statistics for the three mechanism datasets used in Section~\ref{sec:mechanisms}. The implemented columns cover observations whose matched unit falls within the stated buffer of an implemented LSLA deal. The failed columns cover units matched to a failed deal with no deal-specific evidence that community resistance caused the failure. Panel~A is an EA $\times$ round panel for Afrobarometer Rounds 4--9. Panel~B is a deal $\times$ year panel for GDELT GKG over 2015--2025. Panel~C is a constituency $\times$ election panel for CLEA. Afrobarometer missing codes (9, 95, 99) are treated as missing. GDELT outcomes are the inverse hyperbolic sine of annual GKG article counts. Vote shares and turnout are proportions in $[0,1]$.
\end{tablenotes}
\end{threeparttable}
\end{table}

%% file: tables/02_summary_stats.tex
\begin{table}[tbp]
\centering
\begin{threeparttable}
\setlength{\tabcolsep}{10pt}
\caption{Summary Statistics of Implemented and Failed/Abandoned LSLA Deals}
\label{tab:summary_stats}
\begin{tabular}{lcc}
\toprule
\textit{Deal status:} & Implemented & Failed/Abandoned \\
\cmidrule(lr){2-2}\cmidrule(lr){3-3}
 & ($N=1,107$) & ($N=284$) \\
\midrule
\textbf{Panel A: Deal characteristics} \\[3pt]
Intended size (ha), mean & 28,245 & 52,652 \\
Intended size (ha), median & 3,600 & 11,950 \\
Transnational investor (\%) & 61.0 & 84.2 \\
Food crops (\%) & 52.5 & 51.1 \\
Biofuels / biomass (\%) & 8.3 & 34.5 \\
Community land (\%) & 15.4 & 14.4 \\
State land (\%) & 14.3 & 7.4 \\
Private land (\%) & 13.3 & 4.6 \\
Lease (\%) & 47.3 & 56.0 \\
Concession (\%) & 13.1 & 3.9 \\
\midrule
\textbf{Panel B: Geographic distribution} \\[3pt]
East Africa (\%) & 46.6 & 57.0 \\
West Africa (\%) & 31.1 & 30.3 \\
Central Africa (\%) & 14.2 & 2.5 \\
Southern Africa (\%) & 5.1 & 4.9 \\
North Africa (\%) & 3.0 & 5.3 \\
Countries & 37 & 31 \\
\midrule
\textbf{Panel C: Timing} \\[3pt]
Negotiation start, mean year & 2006.9 & 2009.1 \\
Negotiation start, median year & 2009 & 2009 \\
\midrule
\textbf{Panel D: Pre-treatment balance} \\[3pt]
Protests within 50km (5yr avg) & 0.70 & 0.24 \\
All conflict events within 50km (5yr avg) & 2.61 & 1.10 \\
Distance to capital (km) & 343 & 368 \\
Population within 50km (GHSL, 000s) & 944 & 680 \\
Built-up area within 50km (km$^2$, GHSL) & 39.1 & 29.0 \\
\bottomrule
\end{tabular}
\begin{tablenotes}[flushleft]
\footnotesize
\item[] \textit{Notes.} Summary statistics for the analytical sample used in the event study. The treated group consists of implemented deals and the control group consists of failed or abandoned deals whose stated failure reason is unrelated to community resistance. Intended size uses the \textit{intended size in ha} field from the Land Matrix; where missing, the general \textit{deal size} field is used. Investor scope, intended use, prior land ownership, and deal structure are coded from Land Matrix structured fields and are not mutually exclusive. Geographic distribution is based on the country of the deal location, using the UN sub-regional classification for Africa. Panel~D reports pre-treatment characteristics: conflict averages are computed from ACLED events in the five years before each deal's negotiation year; population and built-up surface area are from the Global Human Settlement Layer (GHSL R2023A), extracted within a 50km buffer using the epoch nearest to each deal's negotiation year. None of the Panel~D characteristics significantly predicts deal failure (see Table~\ref{tab:failure_prediction}).
\end{tablenotes}
\end{threeparttable}
\end{table}

%% file: sections/IV._Causal_Evidence.tex
\section{Causal Evidence}
\label{sec:causal_evidence}

\subsection{Main Results}

\begin{figure}[t]
    \centering
    \includegraphics[width=0.7\textwidth]{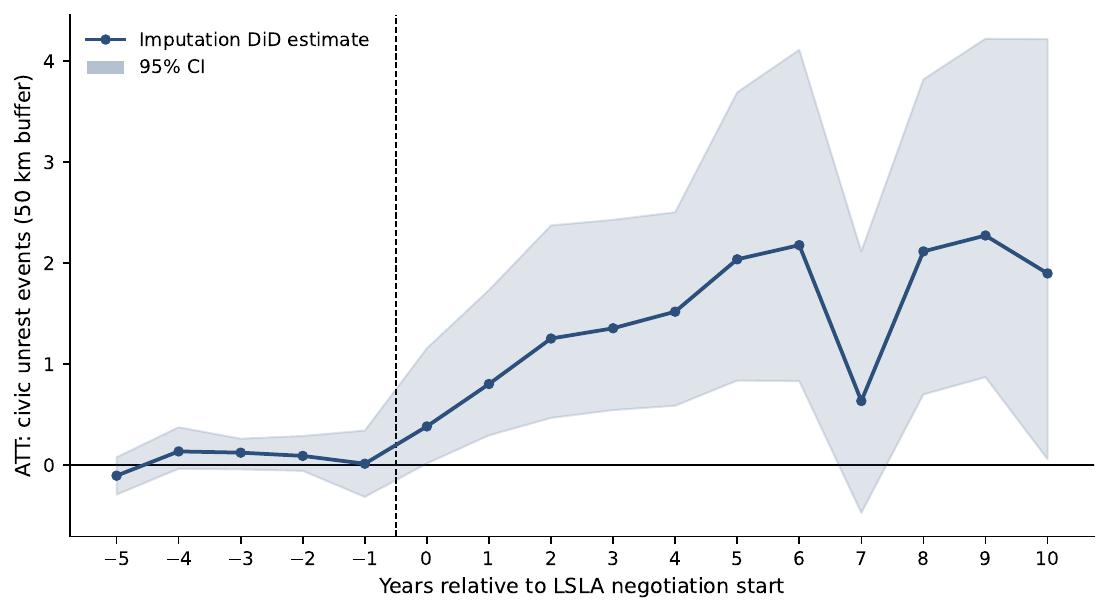}
    \caption{The Effect of LSLAs on Civic Protest Activity}
    \label{fig:main_event_study}
    \caption*{\textit{Notes.} Coefficients obtained via the \citet{borusyak2024revisiting} imputation estimator. The outcome is the annual count of protest and riot events within a $50\,\text{km}$ buffer. The baseline comparison sample consists of implemented deals and failed or abandoned deals with exogenous failure reasons; the imputation uses all untreated observations, including pre-treatment observations of eventually treated units. 95\% confidence intervals are clustered at the deal level.}
\end{figure}

Figure \ref{fig:main_event_study} plots the dynamic event-study coefficients for civic protest activity within the $50\,\text{km}$ buffer surrounding implemented deals, where $t=0$ denotes the year of deal negotiation. Pre-treatment coefficients are close to zero and show no discernible trend prior to implementation, supporting the parallel trends assumption. Following deal completion, I observe a sustained and statistically significant increase in protest activity that persists throughout the decade-long event window, indicating a durable mobilization response.

Table \ref{tab:two_panel_results} presents the aggregate post-treatment ATT estimates under the imputation design. The civic unrest estimates are positive and precisely estimated in the country-by-year specification, and the broader all-conflict response is also positive. The pre-treatment $F$-statistics remain far from conventional significance levels in the preferred specification (column 3), supporting the parallel-trends interpretation.\footnote{Throughout the paper, the pre-trend $F$ reported in the table notes is the average squared $t$-statistic of the pre-treatment event-time leads, with the associated joint $p$-value in brackets. Standard errors are clustered at the deal level using 200 bootstrap iterations, except for leave-one-country-out estimates, which use 100 iterations for computational tractability. Multi-way clustering at both the deal and country-year level yields virtually identical precision (Appendix Section~\ref{sec:appendix_clustering}).}

The civic unrest ATT of approximately 1.48 more than doubles the pre-treatment mean of 0.94 events per year and is roughly four to five times the benchmark rainfall-shock effect of \citet{miguel2004economic}. While standard weather shocks introduce transient income fluctuations, the persistent magnitude of this protest response suggests that the permanent (in the sample period) enclosure of customary land acts as a more severe structural resource shock \citep[as in][]{berman2017mine}, provoking sustained civic mobilization.

Furthermore, this escalation extends well beyond targeted protests. As shown in columns (4) through (6), the point estimates for all conflict events exhibit a similar structural break. The preferred ATT of 3.08 represents a 79\% increase over the baseline mean of 3.90 events per year. Combined with the decade-long persistence shown in the event study, these estimates reveal that large-scale land acquisitions do not generate temporary local friction: they durably destabilize the surrounding security environment.

\input{tables/03_baseline}

\paragraph{Spatial spillovers.} With a $50\,\text{km}$ buffer around each deal, treated catchments can overlap when deals are clustered in the same region, raising a potential concern about the Stable Unit Treatment Value Assumption (SUTVA). At the $50\,\text{km}$ baseline, 92.1\% of treated deals have at least one neighboring treated deal within twice the buffer radius (Table~\ref{tab:buffer_overlap}), reflecting the geographic concentration of agricultural investment zones. This overlap means that some conflict events are counted toward multiple deals simultaneously, which could inflate standard errors but does not bias the point estimate provided that treatment effects are additive. I address this concern directly below by showing that the result is robust to excluding the most heavily overlapping deals. I proceed with the $50\,\text{km}$ buffer as the primary specification; the buffer-sensitivity exercise in this subsection confirms that the results are not sensitive to buffer radius choice.

\subsection{Robustness}
\label{sec:robustness}

\paragraph{Spatial buffers and treatment timing.}
Table \ref{tab:sensitivity_specs} evaluates sensitivity to the spatial definition of the treated area and to the date used to mark treatment. Across the wider buffer sweep reported in Appendix Table \ref{tab:buffer_sweep_master}, protest effects rise monotonically in buffer radius from $25\,\text{km}$ to $100\,\text{km}$, consistent with a genuine treatment effect measured over a broader local neighborhood and inconsistent with specification search. Column (3) in Table \ref{tab:sensitivity_specs} uses the recorded implementation onset year as treatment timing instead of the negotiation start. In the final analytical sample, those implementation-onset estimates are slightly attenuated and less stable on the smaller subsample with recorded implementation dates. 

Table~\ref{tab:later_treated_parallel} then holds the preferred $50\,\text{km}$ imputation specification fixed and changes only the control group. Columns~(1) and~(3) report the baseline \citet{greenstone2010agglomeration} design, which compares implemented deals to failed deals with no deal-specific evidence of community-resistance failure. Columns~(2) and~(4) instead restrict the sample to eventually implemented deals and compare each treated cohort to deals implemented five to ten years later, using only those later cohorts' not-yet-treated observations. This later-treated design therefore keeps the same estimator, outcome definition, timing convention, fixed effects, and controls, while replacing the failed-deal counterfactual with a timing-based comparison drawn entirely from successful projects.

The resulting estimates provide a reassuring robustness check. For civic unrest, the later-treated estimate is slightly smaller than the baseline failed-deal estimate ($1.363$ versus $1.438$) and retains a clean pre-trend diagnostic ($p = 0.17$). For the broader all-conflict outcome, the later-treated estimate is also smaller than the failed-deal estimate ($2.243$ versus $2.915$) but remains positive, precisely estimated, and free of detectable pre-trends. The mild attenuation under the later-treated design is consistent with the failed-deal counterfactual being a slightly more conservative baseline, but does not threaten the qualitative result. The headline protest result therefore survives under a substantially different comparison logic that discards failed deals entirely. Appendix Section~\ref{sec:appendix_later_treated} reports the full design details. Appendix Section~\ref{sec:appendix_spatial_accuracy_restriction} shows that the same ACLED patterns remain visible when the sample is restricted to the high-spatial-accuracy subset.

\input{tables/04_sensitivity_specs}
\input{tables/05_later_treated_parallel}

\paragraph{SUTVA robustness.}
As noted above, 92.1\% of treated deals have at least one overlapping buffer neighbor. To verify that spatial interference across treated units does not drive the main result, Table~\ref{tab:drop_clustered} re-estimates the baseline ATT after progressively dropping treated deals with many nearby treated neighbors. Spatial interference among treated deals does not explain the main finding; even conservative trimming rules leave a sizeable mobilization response.

\begin{figure}[t]
    \centering
    \includegraphics[width=0.5\textwidth]{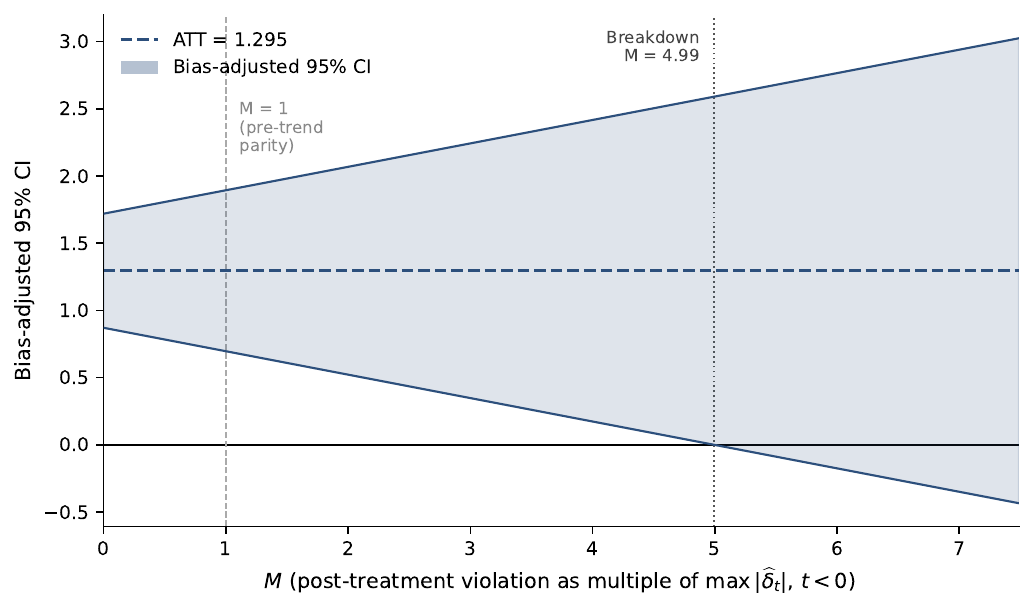}
    \caption{Sensitivity to Violations of Parallel Trends}
    \label{fig:sensitivity}
    \caption*{\textit{Notes.} Confidence intervals from the \citet{rambachan2023credible} sensitivity analysis. Each band shows the range of treatment effects consistent with violations of parallel trends of size at most $M$ times the largest pre-treatment deviation. Outcome is civic protests within a $50\,\text{km}$ buffer.}
\end{figure}

\paragraph{Pre-trend validation.}
I evaluate the robustness of the parallel trends assumption using the bounds approach of \citet{rambachan2023credible}. Figure \ref{fig:sensitivity} shows that the protest effect maintains statistical significance under substantial violations of parallel trends, well beyond the $M = 1$ benchmark where post-treatment violations equal the observed pre-trend evidence. The robustness to sizeable departures from parallel trends is particularly reassuring given that the control group is defined by deal failure instead of true random assignment.

\subsection{Heterogeneity}
\label{sec:heterogeneity_results}

While the baseline estimates reveal an aggregate escalation in civic unrest following land acquisitions, this average treatment effect likely masks important variation in the nature of these deals. The theoretical framework suggests that the intensity of local resistance should depend on who acquires the land, what they intend to do with it, and the institutional status of the territory being enclosed. In this section, I unpack the mechanisms driving the aggregate unrest by exploring treatment heterogeneity along three dimensions: investor origin, crop type composition, and prior land ownership. Together, these tests reveal that the increase in mobilization is not a uniform response to agricultural commercialization, but appears to be concentrated in deals where domestic entities enclose customary land for food crops.

\begin{figure}[t]
    \centering
        \includegraphics[width=0.495\textwidth]{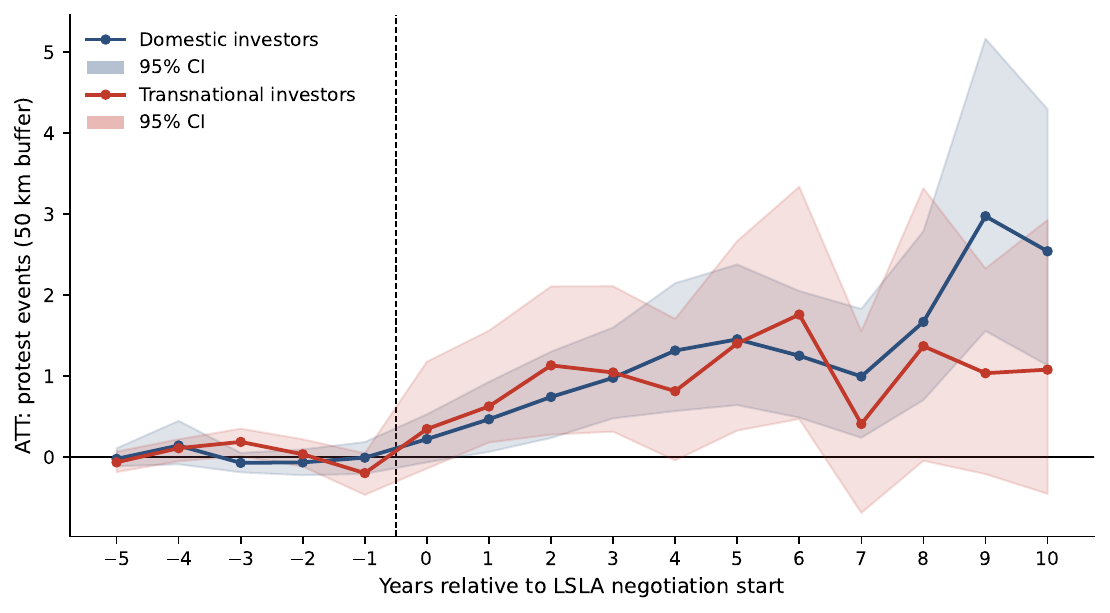}
    \hfill
        \includegraphics[width=0.495\textwidth]{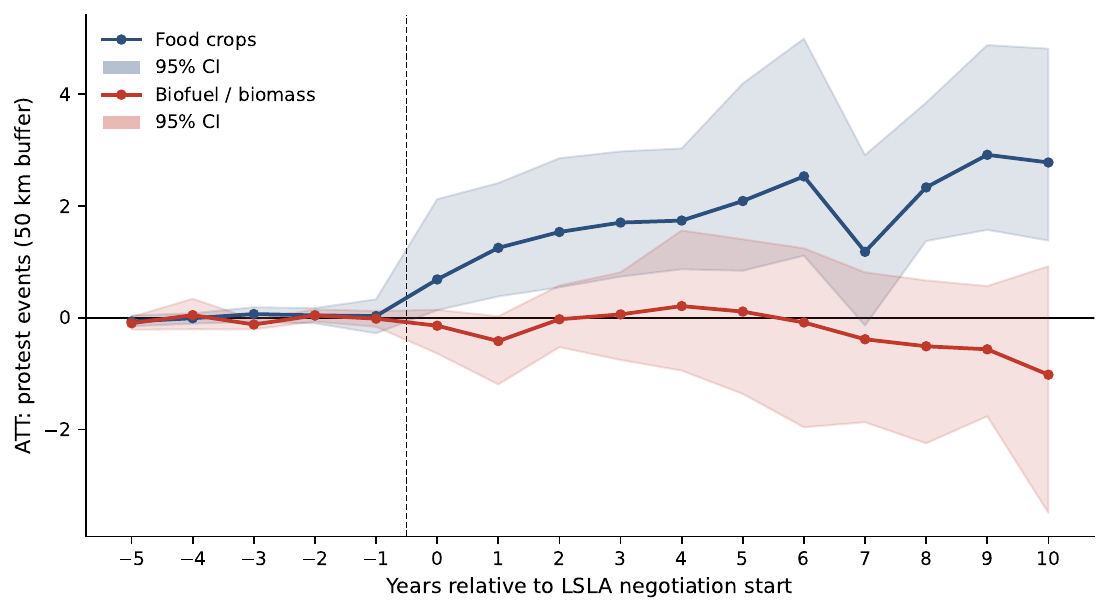}
    \caption{Heterogeneity by Investor Origin and Crop Type}
    \label{fig:heterogeneity_asset}
    \caption*{\textit{Notes.} Failed-control event-study coefficients from the \citet{borusyak2024revisiting} imputation estimator, estimated separately by investor origin (left panel) and intended crop type (right panel). 95\% confidence intervals are clustered at the deal level.}
\end{figure}

\paragraph{Investor origin and crop composition.}
Figure \ref{fig:heterogeneity_asset} (left panel) shows that the escalation in civic unrest is most pronounced among deals led by domestic investors. The transnational-investor estimates are also positive in the final analytical sample, but they are less precise, remain smaller, and are less clearly persistent in the event-study specification. These dynamics point in the direction of the second prediction from Section \ref{sec:background}: domestic actors may be more likely to acquire customary community land through statutory formalization, with greater potential to displace subsistence livelihoods, while facing fewer international ESG constraints that would otherwise mandate community consultation. Biofuel or other non-food deals do not display the same positive effects. Both of these patterns are consistent with the findings in \citet{balestri2021land}.

\paragraph{Land ownership.}
I further examine heterogeneity by land ownership. Since prior-owner information is much sparser among failed controls than among eventually implemented deals,\footnote{When restricting the sample to deals with clear community/indigenous or state prior land ownership, fewer than 20 failed deals remain.} the formal land-ownership comparison is reported separately in Table~\ref{tab:ownership_later_treated_main} using the later-treated design to preserve statistical power.

\input{tables/06_ownership_later_treated_main}

The concentration of the effect in domestic-led, food-crop deals on community and state land provides an empirical indication for the political economy of African state-building articulated by \citet{mamdani1996citizen}, and later by \citet{boone2014property}. The community/indigenous-land panel A shows large and consistent positive effects across both ACLED outcomes, whereas the state-owned land panel B shows a robust protest effect but a less robust all-conflict effect once country-by-year fixed effects and rainfall controls are jointly imposed. The qualitative ownership pattern is therefore strongest for community-titled land, where the estimates are roughly 2.5 times larger than the later-treated civic unrest results in Table~\ref{tab:later_treated_parallel}. When a domestic entity executes a land contract on customary or state-administered land, it is rarely an arm's-length market transaction, as it frequently involves state facilitation within the ambiguous architecture of dual tenure \citep{hall2011land}. Domestic elites face lower international scrutiny and possess the political capital to navigate the opaque dual-tenure system, executing top-down formalizations that bypass community consent \citep{molebatsi2019land,buur2023land}. Foreign actors, operating under more intense international scrutiny and ESG requirements, still generate positive protest responses in the final analytical sample, but the domestic pattern remains more pronounced. The protest effect thus appears strongest where politically connected domestic actors exploit institutional ambiguity.

%% file: tables/03_baseline.tex
\begin{table}[tbp]
\centering
\caption{Effect of Large-Scale Land Acquisitions on Local Conflict}
\label{tab:two_panel_results}
\begin{threeparttable}
\setlength{\tabcolsep}{10pt}
\begin{tabular}{lD{.}{.}{3}D{.}{.}{3}D{.}{.}{3}D{.}{.}{3}D{.}{.}{3}D{.}{.}{3}}
\toprule
\textit{Dependent Variable:} & \multicolumn{3}{c}{Protests} & \multicolumn{3}{c}{All conflict events} \\
\cmidrule(lr){2-4} \cmidrule(lr){5-7}
 & \multicolumn{1}{c}{(1)} & \multicolumn{1}{c}{(2)} & \multicolumn{1}{c}{(3)} & \multicolumn{1}{c}{(4)} & \multicolumn{1}{c}{(5)} & \multicolumn{1}{c}{(6)} \\
\midrule
ATT (post-treatment) & 1.106^{***} & 1.438^{***} & 1.484^{***} & 2.904^{***} & 2.915^{***} & 3.080^{***} \\
 & (0.207) & (0.192) & (0.200) & (0.334) & (0.351) & (0.372) \\
 & [0.640] & [0.587] & [0.624] & [1.560] & [1.297] & [1.336] \\
\addlinespace[4pt]
Pre-trend $F$ & 1.89 & 1.33 & 1.38 & 0.64 & 0.66 & 0.66 \\
 & [0.09] & [0.25] & [0.23] & [0.67] & [0.65] & [0.65] \\
\midrule
Pre-treatment mean & \multicolumn{1}{c}{0.94} & \multicolumn{1}{c}{0.94} & \multicolumn{1}{c}{0.94} & \multicolumn{1}{c}{3.90} & \multicolumn{1}{c}{3.90} & \multicolumn{1}{c}{3.90} \\
$N$ (treated) & \multicolumn{1}{c}{1,107} & \multicolumn{1}{c}{1,107} & \multicolumn{1}{c}{1,107} & \multicolumn{1}{c}{1,107} & \multicolumn{1}{c}{1,107} & \multicolumn{1}{c}{1,107} \\
$N$ (control) & \multicolumn{1}{c}{284} & \multicolumn{1}{c}{284} & \multicolumn{1}{c}{284} & \multicolumn{1}{c}{284} & \multicolumn{1}{c}{284} & \multicolumn{1}{c}{284} \\
\midrule
Deal FE & \multicolumn{1}{c}{\checkmark} & \multicolumn{1}{c}{\checkmark} & \multicolumn{1}{c}{\checkmark} & \multicolumn{1}{c}{\checkmark} & \multicolumn{1}{c}{\checkmark} & \multicolumn{1}{c}{\checkmark} \\
Year FE & \multicolumn{1}{c}{\checkmark} &  &  & \multicolumn{1}{c}{\checkmark} &  &  \\
Country $\times$ Year FE &  & \multicolumn{1}{c}{\checkmark} & \multicolumn{1}{c}{\checkmark} &  & \multicolumn{1}{c}{\checkmark} & \multicolumn{1}{c}{\checkmark} \\
Rainfall control &  &  & \multicolumn{1}{c}{\checkmark} &  &  & \multicolumn{1}{c}{\checkmark} \\
\bottomrule
\end{tabular}
\begin{tablenotes}[para,flushleft]
\footnotesize
\item \textit{Notes.} Each column is a separate event study using the \citet{borusyak2024revisiting} imputation estimator. The unit of observation is an LSLA deal, with outcomes aggregated within a 50\,km buffer. The baseline comparison sample consists of implemented deals ($N=1,107$) and failed deals ($N=284$); the imputation uses all untreated observations, including pre-treatment observations of eventually treated units. The protests outcome counts ACLED ``Protests'' and ``Riots'' events within the buffer. The all-conflict outcome counts all ACLED event types (Battles, Protests, Riots, Violence against civilians, Explosions/Remote violence, and Strategic developments). Bootstrapped standard errors (200 iterations) clustered by deal are reported in parentheses, and \citet{conley1999gmm} spatial HAC standard errors are reported in brackets (Bartlett kernel, 200\,km bandwidth). ATT is the average post-treatment effect ($k \geq 0$). Pre-trend $F$ is the average squared $t$-statistic for $k < 0$; joint $p$-values are reported in brackets. \\
\item $^{*}p<0.10$, $^{**}p<0.05$, $^{***}p<0.01$
\end{tablenotes}
\end{threeparttable}
\end{table}

%% file: tables/04_sensitivity_specs.tex
\begin{table}[tbp]
\centering
\caption{Effect Sensitivity to Buffer Radius and Treatment Timing}
\label{tab:sensitivity_specs}
\begin{threeparttable}
\setlength{\tabcolsep}{10pt}
\begin{tabular}{lD{.}{.}{3}D{.}{.}{3}D{.}{.}{3}D{.}{.}{3}D{.}{.}{3}D{.}{.}{3}}
\toprule
\textit{Dependent Variable:} & \multicolumn{3}{c}{Protests} & \multicolumn{3}{c}{All conflict events} \\
\cmidrule(lr){2-4} \cmidrule(lr){5-7}
 & \multicolumn{1}{c}{(1)} & \multicolumn{1}{c}{(2)} & \multicolumn{1}{c}{(3)} & \multicolumn{1}{c}{(4)} & \multicolumn{1}{c}{(5)} & \multicolumn{1}{c}{(6)} \\
\midrule
ATT (post-treatment) & 0.600^{***} & 1.954^{***} & 0.980^{***} & 0.996^{***} & 4.538^{***} & 2.609^{***} \\
 & (0.116) & (0.237) & (0.262) & (0.178) & (0.706) & (0.684) \\
 & [0.316] & [0.842] & [0.683] & [0.717] & [2.500] & [1.701] \\
\addlinespace[4pt]
Pre-trend $F$ & 1.43 & 1.92 & 1.76 & 0.80 & 0.38 & 0.30 \\
 & [0.21] & [0.09] & [0.12] & [0.55] & [0.86] & [0.92] \\
\midrule
Pre-treatment mean & \multicolumn{1}{c}{0.36} & \multicolumn{1}{c}{1.78} & \multicolumn{1}{c}{1.23} & \multicolumn{1}{c}{1.33} & \multicolumn{1}{c}{7.66} & \multicolumn{1}{c}{4.04} \\
$N$ (treated) & \multicolumn{1}{c}{1,107} & \multicolumn{1}{c}{1,107} & \multicolumn{1}{c}{949} & \multicolumn{1}{c}{1,107} & \multicolumn{1}{c}{1,107} & \multicolumn{1}{c}{949} \\
$N$ (control) & \multicolumn{1}{c}{284} & \multicolumn{1}{c}{284} & \multicolumn{1}{c}{284} & \multicolumn{1}{c}{284} & \multicolumn{1}{c}{284} & \multicolumn{1}{c}{284} \\
\midrule
Deal FE & \multicolumn{1}{c}{\checkmark} & \multicolumn{1}{c}{\checkmark} & \multicolumn{1}{c}{\checkmark} & \multicolumn{1}{c}{\checkmark} & \multicolumn{1}{c}{\checkmark} & \multicolumn{1}{c}{\checkmark} \\
Country $\times$ Year FE & \multicolumn{1}{c}{\checkmark} & \multicolumn{1}{c}{\checkmark} & \multicolumn{1}{c}{\checkmark} & \multicolumn{1}{c}{\checkmark} & \multicolumn{1}{c}{\checkmark} & \multicolumn{1}{c}{\checkmark} \\
\midrule
Buffer 25\,km & \multicolumn{1}{c}{\checkmark} &  &  & \multicolumn{1}{c}{\checkmark} &  &  \\
Buffer 75\,km &  & \multicolumn{1}{c}{\checkmark} &  &  & \multicolumn{1}{c}{\checkmark} &  \\
Buffer 50\,km &  &  & \multicolumn{1}{c}{\checkmark} &  &  & \multicolumn{1}{c}{\checkmark} \\
Implementation onset &  &  & \multicolumn{1}{c}{\checkmark} &  &  & \multicolumn{1}{c}{\checkmark} \\
\bottomrule
\end{tabular}
\begin{tablenotes}[para,flushleft]
\footnotesize
\item \textit{Notes.} Each column is a separate event study. The unit of observation is an LSLA deal with outcomes aggregated within a spatial buffer. Columns (1) and (4) use a 25\,km buffer; columns (2) and (5) use a 75\,km buffer; columns (3) and (6) use the 50\,km buffer with implementation year as the treatment onset instead of negotiation year. All columns use the \citet{borusyak2024revisiting} imputation estimator with country $\times$ year fixed effects. Bootstrapped standard errors (200 iterations) are clustered by deal in parentheses, and \citet{conley1999gmm} spatial HAC standard errors are reported in brackets (Bartlett kernel, 200\,km bandwidth). Pre-trend $F$ is the average squared $t$-statistic for $k < 0$; joint $p$-values are reported in brackets. \\
\item $^{*}p<0.10$, $^{**}p<0.05$, $^{***}p<0.01$
\end{tablenotes}
\end{threeparttable}
\end{table}

%% file: tables/05_later_treated_parallel.tex
\begin{table}[tbp]
\centering
\caption{Later-Treated Timing Robustness at $50\,\text{km}$}
\label{tab:later_treated_parallel}
\begin{threeparttable}
\setlength{\tabcolsep}{8pt}
\begin{tabular}{lcccc}
\toprule
 & \multicolumn{2}{c}{Protests} & \multicolumn{2}{c}{All conflict events} \\
\cmidrule(lr){2-3} \cmidrule(lr){4-5}
 & \shortstack{(1)\\Failed-deal\\controls} & \shortstack{(2)\\Later-treated\\controls} & \shortstack{(3)\\Failed-deal\\controls} & \shortstack{(4)\\Later-treated\\controls} \\
\midrule
ATT (post-treatment) & $1.438^{***}$ & $1.363^{***}$ & $2.915^{***}$ & $2.243^{***}$ \\
 & (0.192) & (0.181) & (0.351) & (0.287) \\
\addlinespace[4pt]
Pre-trend $F$ & 1.65 & 1.62 & 0.73 & 1.79 \\
 & [0.16] & [0.17] & [0.57] & [0.13] \\
\midrule
Pre-treatment mean & 0.94 & 0.85 & 3.90 & 2.71 \\
Implemented deals & 1,107 & 908 & 1,107 & 908 \\
Failed controls & 284 & --- & 284 & --- \\
Cohort stacks & --- & 28 & --- & 28 \\
\midrule
Deal FE & \checkmark & \checkmark & \checkmark & \checkmark \\
Country $\times$ Year FE & \checkmark & \checkmark & \checkmark & \checkmark \\
Rainfall control & \checkmark & \checkmark & \checkmark & \checkmark \\
Treatment onset & Negotiation & Negotiation & Negotiation & Negotiation \\
Control lag window & --- & 5--10 years & --- & 5--10 years \\
\bottomrule
\end{tabular}
\begin{tablenotes}[para,flushleft]
\footnotesize
\item \textit{Notes.} All columns use the \citet{borusyak2024revisiting} imputation estimator on cumulative ACLED outcomes aggregated within a $50\,\text{km}$ buffer, with deal fixed effects, country $\times$ year fixed effects, negotiation-year treatment onset, rainfall controls, and 200 bootstrap iterations clustered at the deal level. Columns (1) and (3) report the preferred implemented-versus-failed Donaldson design from the main text. Columns (2) and (4) hold that estimator and outcome construction fixed but replace the failed-deal counterfactual with a stacked later-treated comparison in which each treated cohort is compared to deals implemented 5 to 10 years later, using only those later cohorts' not-yet-treated observations. The later-treated sample therefore contains only eventually implemented deals; the same deal can serve as a control for earlier cohorts and as treated for its own cohort. ATT is the average post-treatment effect for event times $k \geq 0$. Pre-trend $F$ is the average squared $t$-statistic for pre-treatment leads; joint $p$-values are reported in brackets.
\item $^{*}p<0.10$, $^{**}p<0.05$, $^{***}p<0.01$
\end{tablenotes}
\end{threeparttable}
\end{table}

%% file: tables/06_ownership_later_treated_main.tex
\begin{table}[tbp]
\centering
\caption{Later-Treated Land-Ownership Heterogeneity at $50\,\text{km}$}
\label{tab:ownership_later_treated_main}
\begin{threeparttable}
\setlength{\tabcolsep}{10pt}
\begin{tabular}{lD{.}{.}{3}D{.}{.}{3}D{.}{.}{3}D{.}{.}{3}D{.}{.}{3}D{.}{.}{3}}
\toprule
\textit{Dependent Variable:} & \multicolumn{3}{c}{Protests} & \multicolumn{3}{c}{All conflict events} \\
\cmidrule(lr){2-4} \cmidrule(lr){5-7}
 & \multicolumn{1}{c}{(1)} & \multicolumn{1}{c}{(2)} & \multicolumn{1}{c}{(3)} & \multicolumn{1}{c}{(4)} & \multicolumn{1}{c}{(5)} & \multicolumn{1}{c}{(6)} \\
\midrule
\multicolumn{7}{l}{\textit{Panel A: Community/Indigenous only}} \\
ATT (post-treatment) & 2.497^{***} & 3.550^{***} & 3.621^{***} & 4.620^{***} & 5.238^{***} & 4.783^{***} \\
 & (0.565) & (0.537) & (0.555) & (0.825) & (0.758) & (0.898) \\
 & [1.174] & [1.248] & [1.262] & [1.799] & [1.863] & [1.843] \\
\addlinespace[4pt]
Pre-trend $F$ & 1.44 & 0.92 & 0.79 & 0.62 & 2.05 & 2.02 \\
 & [0.20] & [0.46] & [0.55] & [0.69] & [0.07] & [0.07] \\
\midrule
Pre-treatment mean & \multicolumn{1}{c}{0.40} & \multicolumn{1}{c}{0.40} & \multicolumn{1}{c}{0.40} & \multicolumn{1}{c}{4.99} & \multicolumn{1}{c}{4.99} & \multicolumn{1}{c}{4.99} \\
$N$ (implemented) & \multicolumn{1}{c}{100} & \multicolumn{1}{c}{100} & \multicolumn{1}{c}{100} & \multicolumn{1}{c}{100} & \multicolumn{1}{c}{100} & \multicolumn{1}{c}{100} \\
Cohort stacks & \multicolumn{1}{c}{17} & \multicolumn{1}{c}{17} & \multicolumn{1}{c}{17} & \multicolumn{1}{c}{17} & \multicolumn{1}{c}{17} & \multicolumn{1}{c}{17} \\
\midrule
\multicolumn{7}{l}{\textit{Panel B: State only}} \\
ATT (post-treatment) & 1.805^{***} & 2.040^{***} & 2.114^{***} & 1.265^{*} & 1.457^{***} & 0.866 \\
 & (0.246) & (0.236) & (0.244) & (0.670) & (0.560) & (0.606) \\
 & [0.973] & [1.002] & [1.061] & [1.823] & [1.645] & [1.710] \\
\addlinespace[4pt]
Pre-trend $F$ & 1.81 & 1.97 & 1.92 & 1.61 & 1.66 & 1.69 \\
 & [0.11] & [0.08] & [0.09] & [0.15] & [0.14] & [0.13] \\
\midrule
Pre-treatment mean & \multicolumn{1}{c}{0.37} & \multicolumn{1}{c}{0.37} & \multicolumn{1}{c}{0.37} & \multicolumn{1}{c}{3.44} & \multicolumn{1}{c}{3.44} & \multicolumn{1}{c}{3.44} \\
$N$ (implemented) & \multicolumn{1}{c}{125} & \multicolumn{1}{c}{125} & \multicolumn{1}{c}{125} & \multicolumn{1}{c}{125} & \multicolumn{1}{c}{125} & \multicolumn{1}{c}{125} \\
Cohort stacks & \multicolumn{1}{c}{21} & \multicolumn{1}{c}{21} & \multicolumn{1}{c}{21} & \multicolumn{1}{c}{21} & \multicolumn{1}{c}{21} & \multicolumn{1}{c}{21} \\
\midrule
Deal FE & \multicolumn{1}{c}{\checkmark} & \multicolumn{1}{c}{\checkmark} & \multicolumn{1}{c}{\checkmark} & \multicolumn{1}{c}{\checkmark} & \multicolumn{1}{c}{\checkmark} & \multicolumn{1}{c}{\checkmark} \\
Year FE & \multicolumn{1}{c}{\checkmark} &   &   & \multicolumn{1}{c}{\checkmark} &   &   \\
Country $\times$ Year FE &   & \multicolumn{1}{c}{\checkmark} & \multicolumn{1}{c}{\checkmark} &   & \multicolumn{1}{c}{\checkmark} & \multicolumn{1}{c}{\checkmark} \\
Rainfall control &   &   & \multicolumn{1}{c}{\checkmark} &   &   & \multicolumn{1}{c}{\checkmark} \\
\bottomrule
\end{tabular}
\begin{tablenotes}[para,flushleft]
\footnotesize
\item \textit{Notes.} Two-panel later-treated land-ownership heterogeneity. Panel A restricts the sample to deals whose prior owner was coded as Community or Indigenous only; Panel B restricts to State only. Both panels use mutually exclusive ownership bins. All columns use the stacked later-treated design with the \citet{borusyak2024revisiting} imputation estimator on a $50\,\text{km}$ buffer. Treatment begins at negotiation; controls are deals implemented 5 to 10 years later, using only those later cohorts' not-yet-treated observations. Columns (1)--(3) report the protests outcome (ACLED ``Protests'' and ``Riots''); columns (4)--(6) report all ACLED event types. Bootstrapped standard errors (200 iterations) clustered by deal are reported in parentheses; \citet{conley1999gmm} spatial HAC standard errors are reported in brackets (Bartlett kernel, 200\,km bandwidth). The Conley kernel is applied to the stacked later-treated frame, in which the same deal appears as a separate unit across multiple cohort stacks at identical coordinates; the kernel therefore induces within-deal correlation across stacks, effectively clustering at the deal level through the spatial weighting. Pre-trend $F$ is the average squared $t$-statistic across pre-treatment leads; joint $p$-values are reported in brackets. \\
\item $^{*}p<0.10$, $^{**}p<0.05$, $^{***}p<0.01$
\end{tablenotes}
\end{threeparttable}
\end{table}

%% file: sections/V._Mechanisms.tex
\section{Mechanisms}
\label{sec:mechanisms}

To understand why communities mobilize specifically in response to domestic-led deals on community/indigenous land, this section examines three complementary channels: grievance-specific media coverage (GDELT), the erosion of traditional authority (Afrobarometer), and the electoral response in affected constituencies (CLEA).

\subsection{Media Coverage}

To assess whether local grievance escalation is accompanied by a wider public discourse, I examine media coverage near LSLA sites using the GDELT Global Knowledge Graph (GKG) \citep{leetaru2013gdelt}. The GKG indexes news articles globally and assigns each article a set of thematic tags drawn from a standardized ontology, enabling me to measure changes in the volume of coverage devoted to specific grievance-relevant topics (property rights, corruption, and agricultural production) within spatial buffers around deal locations.

I estimate the same \citet{borusyak2024revisiting} imputation DiD specification, replacing the conflict outcome with the inverse hyperbolic sine of the annual GKG article count for each theme. The GKG panel spans 2015--2025 and is used here to track whether land-rights grievances seem to translate into broader media attention near treated deal sites.

\input{tables/07_gdelt_themes_main}

\begin{figure}[t]
    \centering
    \includegraphics[width=1\textwidth]{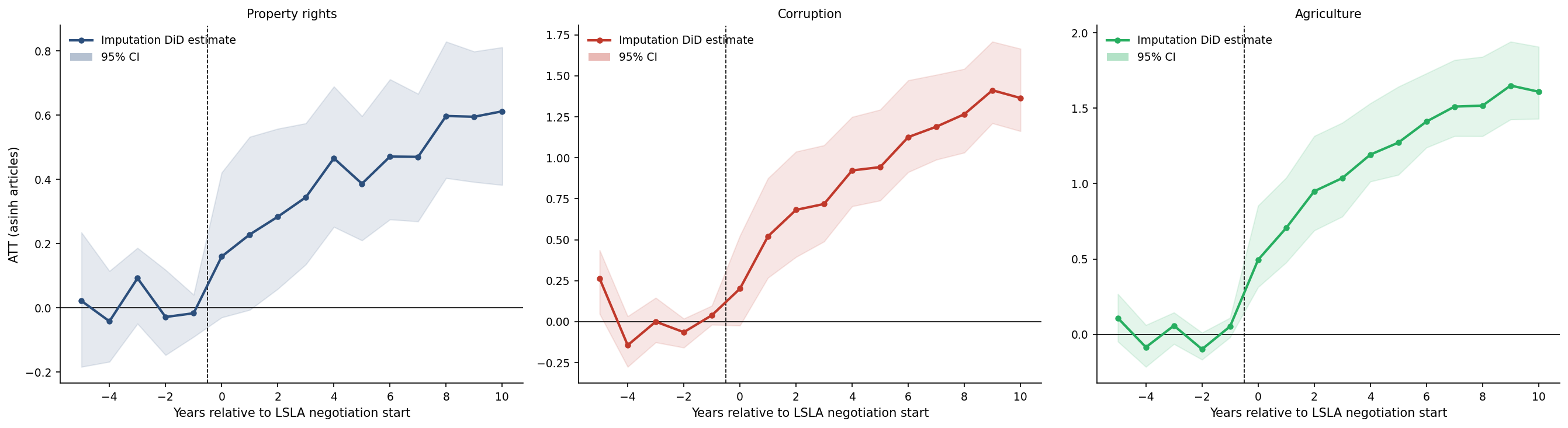}
    \caption{Event-Study Dynamics by Topic of GDELT Thematic Coverage}
    \label{fig:gdelt_themes_es}
    \caption*{\textit{Notes.} Event-study coefficients from the \citet{borusyak2024revisiting} imputation estimator for the inverse hyperbolic sine of the annual GKG article count within a $50\,\text{km}$ buffer, estimated separately by topic: property rights (left panel), corruption (center panel), and agricultural production (right panel). The baseline comparison sample consists of implemented deals and failed deals with no deal-specific evidence that community resistance caused the failure; the imputation uses all untreated observations, including pre-treatment observations of eventually treated units. Country $\times$ Year FE and cluster bootstrap standard errors (200 iterations, clustered at the deal level). GKG v2 panel, 2015--2025. 95\% confidence intervals are clustered at the deal level.}
\end{figure}

Table~\ref{tab:gdelt_themes_main} presents results across the three thematic outcomes. Property-rights, corruption, and agriculture-themed coverage all rise near implemented deals relative to the counterfactual, with a plausible interpretation being a shift in discourse toward land rights, institutional corruption, and agricultural disruption. Media coverage near LSLA sites appears to concentrate in narratives connected to these grievance channels. Theme-level estimates remain robustly positive across the broader buffer sweep $\{25, 50, 75, 100\}\,\text{km}$, with magnitudes rising monotonically in buffer radius (Appendix Table~\ref{tab:buffer_sweep_master}). Figure~\ref{fig:gdelt_themes_es} plots the corresponding event-study dynamics for each theme.

\subsection{Traditional Authority Erosion}
\label{sec:traditional}
Next, I leverage the matched Afrobarometer--deal dataset described in Section~\ref{sec:data}. I first aggregate respondent microdata to an enumeration-area (EA) $\times$ round panel and then estimate the same imputation DiD logic used in the conflict analysis. The sample spans EAs within $50\,\text{km}$ of an identified LSLA (both implemented and failed). The primary specification is:

\begin{equation} \label{eq:afrobarometer}
    \bar{Y}_{c,t} = \beta (\text{Successful Deal}_{c} \times \text{Post}_{c,t}) + \delta_c + \lambda_{q(c),t} + \epsilon_{c,t}
\end{equation}

\noindent where $\bar{Y}_{c,t}$ is the EA-round mean survey outcome for enumeration area $c$ in Afrobarometer round $t$, and $q(c)$ denotes the country containing EA $c$. The displayed equation is the static-DiD analogue presented for intuition; the actual implementation mirrors the conflict analysis and uses the \citet{borusyak2024revisiting} imputation specification, in which $\bar{Y}_{c,t}(0)$ is imputed from untreated EA-round observations using $\delta_c + \lambda_{q(c),t}$ and the EA-level treatment effect $\hat{\tau}_{c,t} = \bar{Y}_{c,t} - \hat{\bar{Y}}_{c,t}(0)$ is then aggregated over post-treatment event time. For the paper's two focal traditional-authority outcomes, contact with a traditional leader is coded as the EA-round share of respondents reporting any contact, while trust in traditional leaders is rescaled from its original 0--3 support to the unit interval before aggregation. I include EA-level fixed effects $\delta_c$ and country-round fixed effects $\lambda_{q(c),t}$, and I cluster standard errors at the EA level. Identification exploits cross-EA variation in deal status (implemented vs.\ failed) within the same $50\,\text{km}$ catchment, mirroring the conflict analysis. Table~\ref{tab:mechanisms_did} reports findings from the imputation DiD specification.

\input{tables/08_mechanisms_did}

Two outcomes remain the clearest survey responses in the sample used here: trust in traditional leaders and contact with traditional leaders to solve community problems. The decline in trust in traditional leaders is economically modest but precisely estimated on the normalized 0--1 scale ($-0.049$, $p<0.01$). It is accompanied by a somewhat larger decline in actual contact with traditional leaders ($-0.083$, $p<0.01$), representing a withdrawal from the primary informal dispute-resolution channel.

The concentration of effects in the customary sphere is theoretically important. The strongest and cleanest survey responses remain in trust in traditional leaders and contact with traditional leaders. This pattern is consistent with custodial failure: communities appear to hold their traditional leaders responsible for the transfer, either because those leaders were perceived as complicit in the deal or because they failed to exercise the protective function that defines their customary role. As these channels lose legitimacy, communities may be left with fewer localized mechanisms through which to redirect grievances away from public protest.

\subsection{Electoral Response}

If LSLAs generate sustained grievances and erode local institutional trust, affected communities may channel these grievances into political opposition. I test this using constituency-level electoral data from the Comparative Legislative Elections Archive \citep{kollman2024clea}, which covers national legislative elections across a broad set of African countries with geocoded constituency boundaries.

I construct a constituency-level panel by assigning each constituency a treatment indicator based on whether an implemented deal falls within a $50\,\text{km}$ buffer of its centroid, using the same spatial scale as the conflict, survey, and media outcomes and following the failed-deal control-group design (failed deals only, with community-resistance cases excluded under the baseline failed-deal screening rule described in Appendix Section~\ref{sec:appendix_failure_hierarchy}). A single paper-wide buffer avoids any appearance of outcome-specific specification search and reflects the fact that communities mobilize, trust erodes, media coverage expands, and constituencies vote at the same community and traditional-authority scale. The $25\,\text{km}$ alternative, which is close to the median radius of legislative constituencies in the sample, is reported alongside the full $\{25, 50, 75, 100\}\,\text{km}$ sweep in Appendix Section~\ref{sec:appendix_buffer_sweep}. I estimate the BJS imputation DiD with constituency fixed effects and Country $\times$ election-year fixed effects.

\input{tables/09_electoral_main}

\begin{figure}[t]
    \centering
    \includegraphics[width=0.49\textwidth]{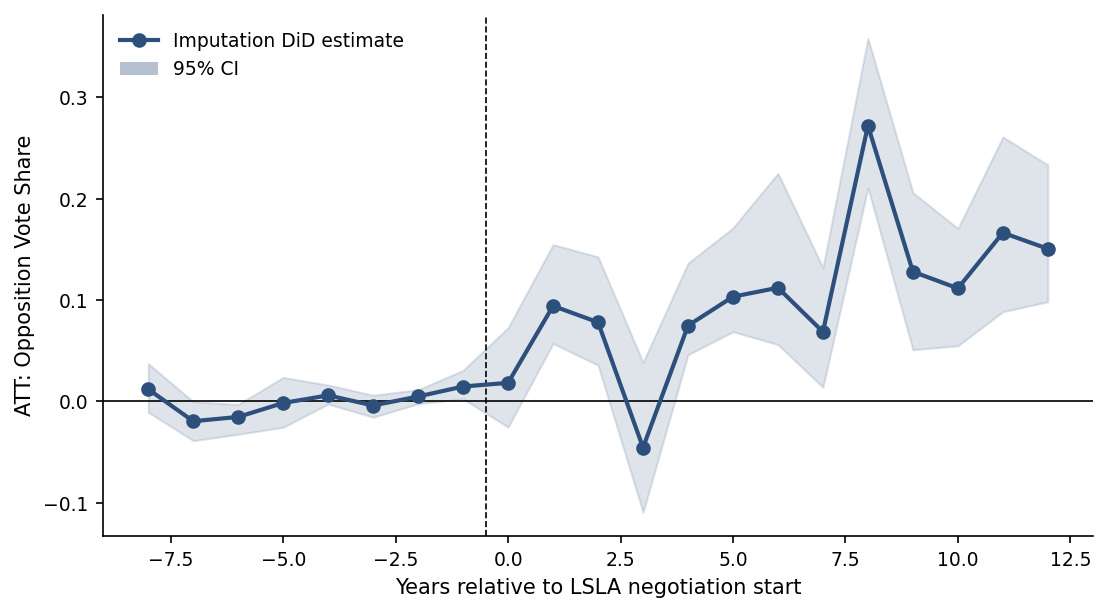}
    \hfill
    \includegraphics[width=0.49\textwidth]{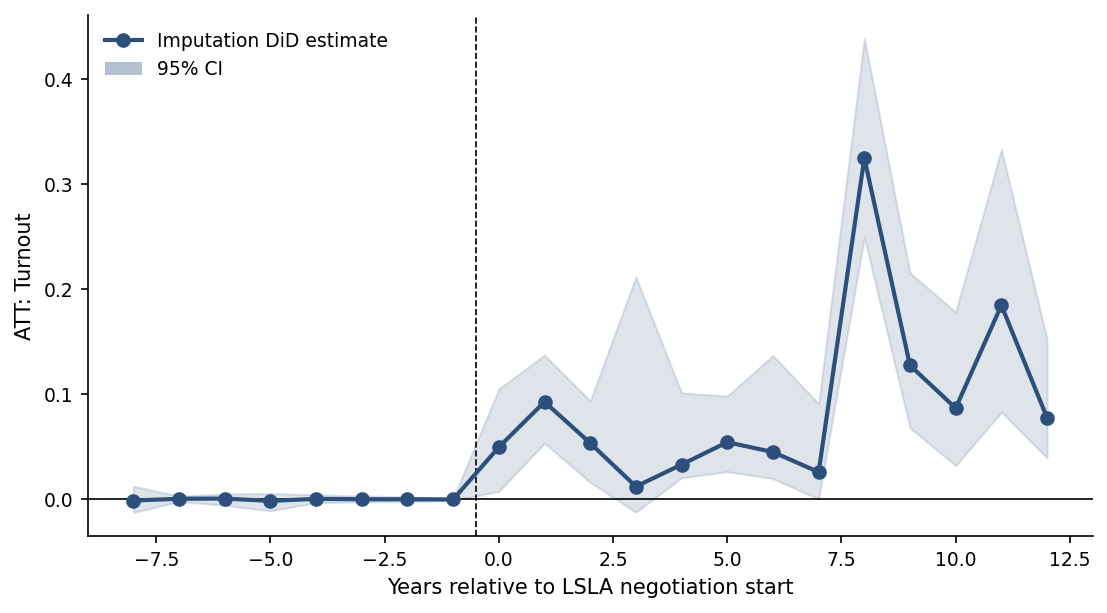}
    \caption{Dynamic Event-Study Estimates of Electoral Response}
    \label{fig:electoral_es}
    \caption*{\textit{Notes.} Event-study coefficients from the \citet{borusyak2024revisiting} imputation estimator for opposition vote share (left panel) and voter turnout (right panel) within a $50\,\text{km}$ buffer of constituency centroids. The baseline comparison sample consists of constituencies within buffer of implemented LSLA deals and constituencies within buffer of failed deals with no deal-specific evidence that community resistance caused the failure; the imputation uses all untreated observations, including pre-treatment observations of eventually treated units. Constituency FE and Country $\times$ election-year FE; cluster bootstrap standard errors (200 iterations, clustered at the constituency level). The dashed vertical line marks $t = 0$ (deal negotiation year). Elections occur every 4--5 years, producing an irregular event-time grid. 95\% confidence intervals are clustered at the constituency level.}
\end{figure}

Table~\ref{tab:electoral_main} presents results for three electoral outcomes. In the preferred Country $\times$ election-year specification, implemented LSLAs increase opposition vote share by 10.0 percentage points and voter turnout by 8.8 percentage points, while the incumbent vote share is essentially unchanged (+0.8 percentage points, not statistically distinguishable from zero). Specifications including only year fixed effects yield a somewhat larger opposition effect (13.2 percentage points) and a modest negative incumbent coefficient ($-2.7$ percentage points), suggesting that absorbing country-by-election-year shocks attenuates the apparent incumbent punishment while leaving the broader mobilization patterns intact. The substantive interpretation is therefore mobilization rather than vote-switching: opposition vote shares and turnout rise together in affected constituencies, with no detectable erosion of the incumbent's vote share under the preferred specification. The pattern is most naturally read as previously non-voting or marginally attached residents being activated against the status quo, rather than incumbent supporters defecting to challengers. Pre-trend tests are clean for turnout but borderline-significant for the preferred opposition and incumbent specifications ($p = 0.05$ and $p = 0.04$ respectively), so these estimates should be interpreted more cautiously than the ACLED protest core. Figure~\ref{fig:electoral_es} plots the corresponding event-study dynamics for opposition vote share and voter turnout.

The three channels operate at different institutional scales and timescales, but together they form a coherent sequence. Traditional authority erosion is the immediate institutional response: the informal dispute-resolution infrastructure that might otherwise contain grievances is weakened at the community level within the survey period. Grievance-specific media escalation follows on an annual cycle, as land-rights and corruption narratives enter public discourse near LSLA sites. Electoral politics represents the downstream response, operating on the 4--5 year electoral cycle, with communities translating institutional grievances into higher turnout and stronger support for opposition parties, even if the incumbent response is no longer cleanly negative in the sample used here. Together, these three channels describe a process in which dispossession moves from private institutional failure to public mobilization to formal political activation.

This sequence also helps explain why the protest effect documented in Section~\ref{sec:causal_evidence} is strongest in domestic-led deals on community and state land. Domestic investors acquire land precisely through the customary and administrative spheres, relying on traditional leaders and state intermediaries to facilitate or acquiesce to transfers. When those institutions fail to secure community consent, the custodial-failure channel is activated directly. Foreign investors operate under different institutional constraints, including ESG accountability requirements and less reliance on politically mediated land acquisition, which is consistent with the smaller and less sharply estimated responses in the heterogeneity analysis. The survey results therefore point most strongly toward an institutionally targeted response within the customary sphere, not a generalized withdrawal from public authority.

%% file: tables/07_gdelt_themes_main.tex
\begin{table}[tbp]
\centering
\caption{GKG Theme Coverage near LSLA Sites}
\label{tab:gdelt_themes_main}
\begin{threeparttable}
\setlength{\tabcolsep}{6pt}
\begin{tabular}{lD{.}{.}{7}D{.}{.}{5}D{.}{.}{5}}
\toprule
\textit{Dependent variable:} & \multicolumn{1}{c}{Property rights} & \multicolumn{1}{c}{Corruption} & \multicolumn{1}{c}{Agriculture} \\
\cmidrule(lr){2-2} \cmidrule(lr){3-3} \cmidrule(lr){4-4}
\multicolumn{1}{l}{\textit{}} & \multicolumn{1}{c}{(1)} & \multicolumn{1}{c}{(2)} & \multicolumn{1}{c}{(3)} \\
\midrule
ATT (post-treatment) & 0.472^{***} & 1.074^{***} & 1.348^{***} \\
 & (0.036) & (0.047) & (0.044) \\
 & [0.086] & [0.152] & [0.181] \\
\addlinespace
Pre-trend $F$ & 0.59 & 2.67 & 2.30 \\
 & [0.70] & [0.02] & [0.04] \\
\midrule
$N$ (treated deals) & \multicolumn{1}{c}{949} & \multicolumn{1}{c}{949} & \multicolumn{1}{c}{949} \\
$N$ (control deals) & \multicolumn{1}{c}{226} & \multicolumn{1}{c}{226} & \multicolumn{1}{c}{226} \\
\midrule
Deal FE & \multicolumn{1}{c}{\checkmark} & \multicolumn{1}{c}{\checkmark} & \multicolumn{1}{c}{\checkmark} \\
Country $\times$ Year FE & \multicolumn{1}{c}{\checkmark} & \multicolumn{1}{c}{\checkmark} & \multicolumn{1}{c}{\checkmark} \\
Rainfall control & \multicolumn{1}{c}{\checkmark} & \multicolumn{1}{c}{\checkmark} & \multicolumn{1}{c}{\checkmark} \\
\bottomrule
\end{tabular}
\begin{tablenotes}[para,flushleft]
\footnotesize
\item \textit{Notes.} Pooled post-treatment ATT from the \citet{borusyak2024revisiting} imputation estimator. The unit of observation is a deal-year, with outcomes measured as the inverse hyperbolic sine of the annual GKG article count within a $50\,\text{km}$ buffer. The baseline comparison sample consists of implemented deals and failed deals with no deal-specific evidence that community resistance caused the failure; the imputation uses all untreated observations, including pre-treatment observations of eventually treated units. Treatment begins at negotiation. All columns include deal fixed effects, Country $\times$ Year FE, and rainfall controls. Standard errors in parentheses come from 200 bootstrap iterations clustered at the deal level; \citet{conley1999gmm} spatial HAC standard errors are reported in brackets. GKG v2 geocoded articles, 2015--2025. Columns (1)--(3) report the thematic outcomes. \\
\item $^{*}p<0.10$, $^{**}p<0.05$, $^{***}p<0.01$
\end{tablenotes}
\end{threeparttable}
\end{table}

%% file: tables/08_mechanisms_did.tex
\begin{table}[tbp]
\centering
\caption{EA-level Effect of LSLAs on Traditional Authority}
\label{tab:mechanisms_did}
\begin{threeparttable}
\setlength{\tabcolsep}{12pt}
\begin{tabular}{lD{.}{.}{8}D{.}{.}{8}}
\toprule
 & \multicolumn{1}{c}{\textit{Contact trad.\ leader}} & \multicolumn{1}{c}{\textit{Trust trad.\ leaders}} \\
\cmidrule(lr){2-2} \cmidrule(lr){3-3}
 & \multicolumn{1}{c}{(1)} & \multicolumn{1}{c}{(2)} \\
\midrule
ATT (post-treatment) & -0.083^{***} & -0.049^{***} \\
 & (0.024) & (0.018) \\
 & [0.043] & [0.023] \\
\addlinespace
Pre-trend $F$ & 0.56 & 0.57 \\
\midrule
$N$ (treated EAs) & \multicolumn{1}{c}{2,240} & \multicolumn{1}{c}{2,240} \\
$N$ (control EAs) & \multicolumn{1}{c}{540} & \multicolumn{1}{c}{540} \\
\midrule
EA FE & \multicolumn{1}{c}{\checkmark} & \multicolumn{1}{c}{\checkmark} \\
Country $\times$ Round FE & \multicolumn{1}{c}{\checkmark} & \multicolumn{1}{c}{\checkmark} \\
\bottomrule
\end{tabular}
\begin{tablenotes}[para,flushleft]
\footnotesize
\item \textit{Notes.} Pooled post-treatment ATT from the \citet{borusyak2024revisiting} imputation estimator. The unit of observation is an Afrobarometer enumeration area (EA), observed in an EA $\times$ round panel from Rounds 4--9. The baseline comparison sample consists of EAs within $50\,\text{km}$ of implemented LSLA deals and EAs within $50\,\text{km}$ of failed deals with no deal-specific evidence that community resistance caused the failure; the imputation uses all untreated observations, including pre-treatment observations of eventually treated units. All columns include EA fixed effects and Country $\times$ Round FE. Bootstrapped standard errors (200 iterations) clustered at the EA level are reported in parentheses, and \citet{conley1999gmm} spatial HAC standard errors are reported in brackets (Bartlett kernel, 200\,km bandwidth, EA centroids). Outcomes are EA-round means built from respondent microdata: column (1) is the share reporting any contact with a traditional leader in the past year, and column (2) rescales trust in traditional leaders from the original 0--3 scale to the unit interval. Pre-trend $F$ is the average squared $t$-statistic across pre-treatment horizons. \\
\item $^{*}p<0.10$, $^{**}p<0.05$, $^{***}p<0.01$
\end{tablenotes}
\end{threeparttable}
\end{table}

%% file: tables/09_electoral_main.tex
\begin{table}[tbp]
\centering
\caption{LSLA Deals and Voting Outcomes at $50\,\text{km}$}
\label{tab:electoral_main}
\begin{threeparttable}
\setlength{\tabcolsep}{4pt}
\begin{tabular}{lD{.}{.}{3}D{.}{.}{5}D{.}{.}{3}D{.}{.}{6}D{.}{.}{3}D{.}{.}{3}}
\toprule
 & \multicolumn{2}{c}{\textit{Incumbent vote share}} & \multicolumn{2}{c}{\textit{Opposition vote share}} & \multicolumn{2}{c}{\textit{Voter turnout}} \\
\cmidrule(lr){2-3} \cmidrule(lr){4-5} \cmidrule(lr){6-7}
 & \multicolumn{1}{c}{(1)} & \multicolumn{1}{c}{(2)} & \multicolumn{1}{c}{(3)} & \multicolumn{1}{c}{(4)} & \multicolumn{1}{c}{(5)} & \multicolumn{1}{c}{(6)} \\
\midrule
ATT (post-treatment) & -0.027^{**} & 0.008 & 0.132^{***} & 0.100^{***} & 0.048^{***} & 0.088^{***} \\
 & (0.012) & (0.010) & (0.016) & (0.009) & (0.007) & (0.009) \\
 & [0.020] & [0.021] & [0.048] & [0.024] & [0.024] & [0.032] \\
\addlinespace
Pre-trend $F$ & 1.73 & 2.00 & 2.58 & 1.97 & 0.12 & 0.07 \\
 & [0.09] & [0.04] & [0.01] & [0.05] & [1.00] & [1.00] \\
\midrule
$N$ (treated constituencies) & \multicolumn{1}{c}{1,536} & \multicolumn{1}{c}{1,536} & \multicolumn{1}{c}{1,536} & \multicolumn{1}{c}{1,536} & \multicolumn{1}{c}{1,536} & \multicolumn{1}{c}{1,536} \\
$N$ (control constituencies) & \multicolumn{1}{c}{146} & \multicolumn{1}{c}{146} & \multicolumn{1}{c}{146} & \multicolumn{1}{c}{146} & \multicolumn{1}{c}{146} & \multicolumn{1}{c}{146} \\
\midrule
Constituency FE & \multicolumn{1}{c}{\checkmark} & \multicolumn{1}{c}{\checkmark} & \multicolumn{1}{c}{\checkmark} & \multicolumn{1}{c}{\checkmark} & \multicolumn{1}{c}{\checkmark} & \multicolumn{1}{c}{\checkmark} \\
Year FE & \multicolumn{1}{c}{\checkmark} &  & \multicolumn{1}{c}{\checkmark} &  & \multicolumn{1}{c}{\checkmark} &  \\
Country $\times$ Year FE &  & \multicolumn{1}{c}{\checkmark} &  & \multicolumn{1}{c}{\checkmark} &  & \multicolumn{1}{c}{\checkmark} \\
\bottomrule
\end{tabular}
\begin{tablenotes}[para,flushleft]
\footnotesize
\item \textit{Notes.} Pooled post-treatment ATT from the \citet{borusyak2024revisiting} imputation estimator in a constituency-level panel from the CLEA electoral database. The baseline comparison sample consists of constituencies within $50\,\text{km}$ of implemented LSLA deals and constituencies within $50\,\text{km}$ of failed deals with no deal-specific evidence that community resistance caused the failure; the imputation uses all untreated observations, including pre-treatment observations of eventually treated units. All columns include constituency fixed effects and either election-year FE (columns 1, 3, and 5) or Country $\times$ election-year FE (columns 2, 4, and 6). Bootstrapped standard errors (200 iterations) clustered at the constituency level are reported in parentheses, and \citet{conley1999gmm} spatial HAC standard errors are reported in brackets (Bartlett kernel, 200\,km bandwidth, constituency centroids). LP-DiD is not reported because elections every 4--5 years leave only 36\% of treated constituencies with any pre-cohort election in the CLEA panel, causing the estimator to drop the majority of units. Pre-trend $F$ is the average squared $t$-statistic for $k < 0$ horizons; joint $p$-values are reported in brackets. \\
\item $^{*}p<0.10$, $^{**}p<0.05$, $^{***}p<0.01$
\end{tablenotes}
\end{threeparttable}
\end{table}

%% file: sections/VI._Conclusion.tex
\section{Conclusion}
\label{sec:conclusion}

This paper provides causal evidence that large-scale land acquisitions in Africa generate sustained civic mobilization, with the strongest effects concentrated among acquisitions of customary land. The results point to local dispossession and domestic political capture as central drivers of rural instability in this setting, more clearly than to a transnational land-grab narrative. The mechanism evidence is suggestive: when domestic elites exploit the ambiguity of dual-tenure systems to acquire customary land with state facilitation, communities appear to respond through civic and electoral channels, and the customary governance structures that might otherwise contain those grievances are weakened in the process.

These findings carry implications for both theory and governance. The evolutionary theory of land rights predicts that rising scarcity drives an organic transition from customary to formal tenure, with state formalization as a welfare-enhancing accelerant. The evidence corroborates that this logic breaks down where formalization is imposed top-down by domestic elites \citep[as suggested by][]{platteau1996evolutionary}: such transfers bypass the endogenous community negotiations the theory envisions, extinguish customary use rights, and generate sustained civic unrest instead of a smooth institutional transition. On the governance side, international frameworks for responsible agricultural investment focus overwhelmingly on foreign sovereigns and agribusinesses; the evidence here suggests domestic elites operating on customary land as the more relevant source of grievance. Three institutional levers follow from the mechanisms: documented community consent should be a precondition for any state-facilitated customary land lease; the fiduciary accountability of traditional leaders as land custodians should be enforceable at the community level; and the ESG-style due diligence requirements currently imposed on transnational investors should be extended to large domestic acquirers, closing a regulatory gap that seems to enable elite capture.

%% file: sections/VII._Appendix.tex
\clearpage

\counterwithin{table}{section}
\counterwithin{figure}{section}
\counterwithin{equation}{section}

\appendix

\section{Data Construction and Control Selection}
\label{sec:appendix_data}

This appendix section expands on the data construction summarized in Section~\ref{sec:data}. It covers (i) the construction of the Land Matrix analytical sample, (ii) the hierarchical classification of deal failure reasons and validation of the control group, (iii) the ACLED outcome aggregation procedure, (iv) the spatial matching of the three mechanism datasets, and (v) the definitions of the investor-origin, land-type, and crop-use classifications used in the heterogeneity analysis.

\subsection{Land Matrix Sample Construction}
\label{sec:appendix_sample_construction}

Appendix Table~\ref{tab:control_sample_waterfall} documents the sample construction from the raw Land Matrix universe to the geocoded ACLED comparison sample. The table first restricts the global Land Matrix download to African-targeted deals and then to the 38 study countries represented in the final analytical sample. It next isolates the study-country deals with retained subnational-precision coordinates and then the implemented-versus-failed comparison sample used in the \citet{greenstone2010agglomeration,donaldson2018railroads}-style design. The bottom rows report the treated branch, the failed-control branch, the endogenous-failure screening applied to failed deals, and the resulting analytical sample of 1,107 implemented deals and 284 failed baseline controls, for a total of 1,391 units.

\subsection{Classification of Failure Reasons}
\label{sec:appendix_failure_hierarchy}

The control group is constructed from abandoned and failed deals using a hierarchical classification of recorded failure evidence. I parse (i) the free-text comments on deal status, (ii) the structured \textit{Presence of land conflicts} field, and (iii) the structured \textit{Community reaction} field. A deal is labeled with community resistance if any of these sources contains an explicit mention of local opposition, resistance, protest activity, or community refusal. I then conduct a supplementary review of previously unresolved failed deals using Land Matrix deal pages, attached source URLs and uploaded-file references, and broader web searches. In the final analytical sample, this review identifies nine additional failed deals with deal-specific evidence that local resistance contributed to the failure. Those cases are excluded from the baseline control group alongside the 58 failures already flagged by the structured/text screening rule.

\input{tables/A1_control_sample_waterfall}

Table~\ref{tab:control_sample_waterfall} therefore serves two purposes. Panel~A documents how the analytical sample is built from the Land Matrix universe to the geocoded ACLED comparison sample. Panel~B shows how the 351 failed or abandoned candidate controls are screened down to the 284 failed baseline controls used in the main specifications. The retained control group consists of failed deals for which the review process finds no deal-specific evidence that community resistance caused the breakdown. Table~\ref{tab:confirmed_exogenous_controls} then reports a deliberately stricter confirmed-exogenous subset as an additional robustness check.

\subsection{Confirmed Exogenous Control Sample}
\label{sec:appendix_confirmed_exogenous_controls}

As a deliberately stricter robustness check, Table~\ref{tab:confirmed_exogenous_controls} restricts the failed-control sample to cases with affirmative evidence of non-community failure. In the final analytical sample, this confirmed-control subset contains 83 failed deals. The treated sample is unchanged. The preferred country-by-year specifications remain positive and statistically significant for civic unrest, although the year-fixed-effect protest column is attenuated and imprecise, as expected in a substantially lower-powered control sample.

\input{tables/A2_confirmed_exogenous_controls}

\subsection{ACLED Outcome Construction}
\label{sec:appendix_acled}

Conflict outcomes are aggregated from ACLED event-level records into annual deal-level counts. For each deal $d$ and year $t$, I compute the count $y^r_{d,t} = \sum_{c \in C_t} \mathbf{1}\{\text{dist}(c, \text{location}_d) \leq r\}$ of events within a buffer of radius $r \in \{25, 50, 75, 100\}\,\text{km}$ around the deal's geocoded coordinates. The \textit{protests} outcome sums ACLED ``Protests'' and ``Riots'' event types; the \textit{all-conflict} outcome sums all ACLED event types (Battles, Protests, Riots, Violence against civilians, Explosions/Remote violence, Strategic developments). Appendix Section~\ref{sec:appendix_armed_conflict} reports a supplementary armed-conflict decomposition in which the composite outcome sums Battles, Explosions/Remote violence, and Violence against civilians, with the component event types, strategic developments, and fatalities shown separately.

\subsection{Matching of the Mechanism Datasets}
\label{sec:appendix_matching}

Each of the three mechanism datasets is matched to the deal panel using a spatial buffer around the deal coordinates. For Afrobarometer, survey enumeration areas are matched to deals within a $50\,\text{km}$ buffer; raw individual-level responses are then collapsed to an EA $\times$ round panel (waves R4--R9), yielding 2{,}240 treated EAs and 540 control EAs. For GDELT GKG, geocoded v2 articles are matched to deals within a $50\,\text{km}$ buffer for the period 2015--2025, and annual article counts enter in inverse-hyperbolic-sine form. For CLEA, legislative-election constituencies are matched to deals within a $50\,\text{km}$ buffer of the constituency centroid, producing a constituency $\times$ election panel. Each matching procedure produces a panel on which the baseline imputation DiD is estimated with deal-level (or EA-level, or constituency-level) fixed effects and country-by-time fixed effects.

\subsection{Classification of Investor Origin, Land Type, and Crop Use}
\label{sec:appendix_deal_types}

Deals are classified as \textit{domestic} if all listed investor parents are nationals of the target country and as \textit{transnational} otherwise. Land type is taken from the Land Matrix \textit{prior land ownership} field. In the ownership-refresh heterogeneity package, the main-text land table uses mutually exclusive \textit{community/indigenous only} and \textit{state only} categories under the later-treated design, while Appendix Section~\ref{sec:appendix_heterogeneity} reports broader mention-based ownership definitions as robustness checks. Crop use is taken from the \textit{intended production} field and collapsed into \textit{food crops} (staples, vegetables, livestock feed) and \textit{biofuels / non-food} (oil palm, sugarcane for ethanol, jatropha, other non-food industrial crops). These classifications are used in the heterogeneity results in Section~\ref{sec:heterogeneity_results}.

\subsection{Pre-treatment Covariates Do Not Predict Failure}
\label{sec:appendix_failure_prediction}

A key identifying assumption is that the control group of failed deals was not systematically selected based on pre-existing local conflict or political conditions. Table~\ref{tab:failure_prediction} provides a narrow diagnostic for this concern by regressing the probability of deal failure on a set of pre-treatment local characteristics: the average annual protest count and total event count within a $50\,\text{km}$ buffer over the five years before the deal's negotiation year, the log distance from the deal site to the country's capital city, and a full set of country fixed effects.

\input{tables/A3_failure_prediction}

The resulting pattern is mixed rather than strongly predictive. Pre-treatment protests and total pre-treatment events are both negative, and geographic distance remains imprecise. The pseudo-$R^{2}$ is still driven largely by country-specific variation, which is consistent with differences in Land Matrix documentation intensity across reporting environments. The appropriate reading is therefore limited: this regression does not show a clean relationship between baseline local conflict conditions and failure status, which helps rule out the most obvious form of endogenous control-group selection. It does not, by itself, establish parallel trends; that evidence comes more directly from the event-study pre-trends and the additional robustness exercises reported in the paper.

\subsection{Dropping Deals with Any Resistance Mention}
\label{sec:appendix_endogenous_failure}

As an additional robustness check, I tighten the control-group filter: in addition to the baseline exclusion rule (which drops failed deals whose primary failure reason is classified as community resistance), I also drop any deal with \emph{any} keyword mention of local resistance, opposition, or protest in the free-text comments or the community-reaction field, even if the primary failure reason is classified differently. Column~(3) of Table~\ref{tab:robustness_master} reports this tightened-filter version alongside the baseline specification. The estimated ATT on protests remains positive, highly significant, and nearly unchanged relative to the baseline ($1.438$ vs.\ $1.351$), confirming that the result is not driven by residual resistance-linked deals in the control group.

\subsection{Composition-Balanced ATT}
\label{sec:appendix_composition}

A second concern about the failed-deals control group is compositional: the treated pool contains 39.0\% domestic investors versus 15.8\% in the control pool. Because the main-text heterogeneity analysis shows that the protest effect is stronger in the domestic subset, an imbalance in investor origin could in principle inflate the pooled ATT. Table~\ref{tab:composition_balanced} reports four diagnostics. The domestic-only subsample remains estimable in the final analytical sample with 45 domestic failed controls, yielding a positive and significant ATT ($1.786$); the transnational-only subsample also yields a positive and significant ATT ($1.396$). Column (4) applies inverse-propensity reweighting that rebalances the control pool to match the treated investor-origin distribution: the reweighted ATT ($1.546$) is slightly larger than the pooled baseline ($1.438$), moving in the opposite direction of a mechanical composition artifact. Column (5) drops the eight controls whose documented failure involves administrative or government action; the ATT remains essentially unchanged at $1.448$. The composition diagnostics therefore continue to suggest that the baseline effect is not an artifact of investor-origin imbalance.

\input{tables/A4_composition_balanced}

\subsection{Leave-One-Country-Out}
\label{sec:appendix_loco}

\begin{figure}[t]
    \centering
    \includegraphics[width=0.85\textwidth]{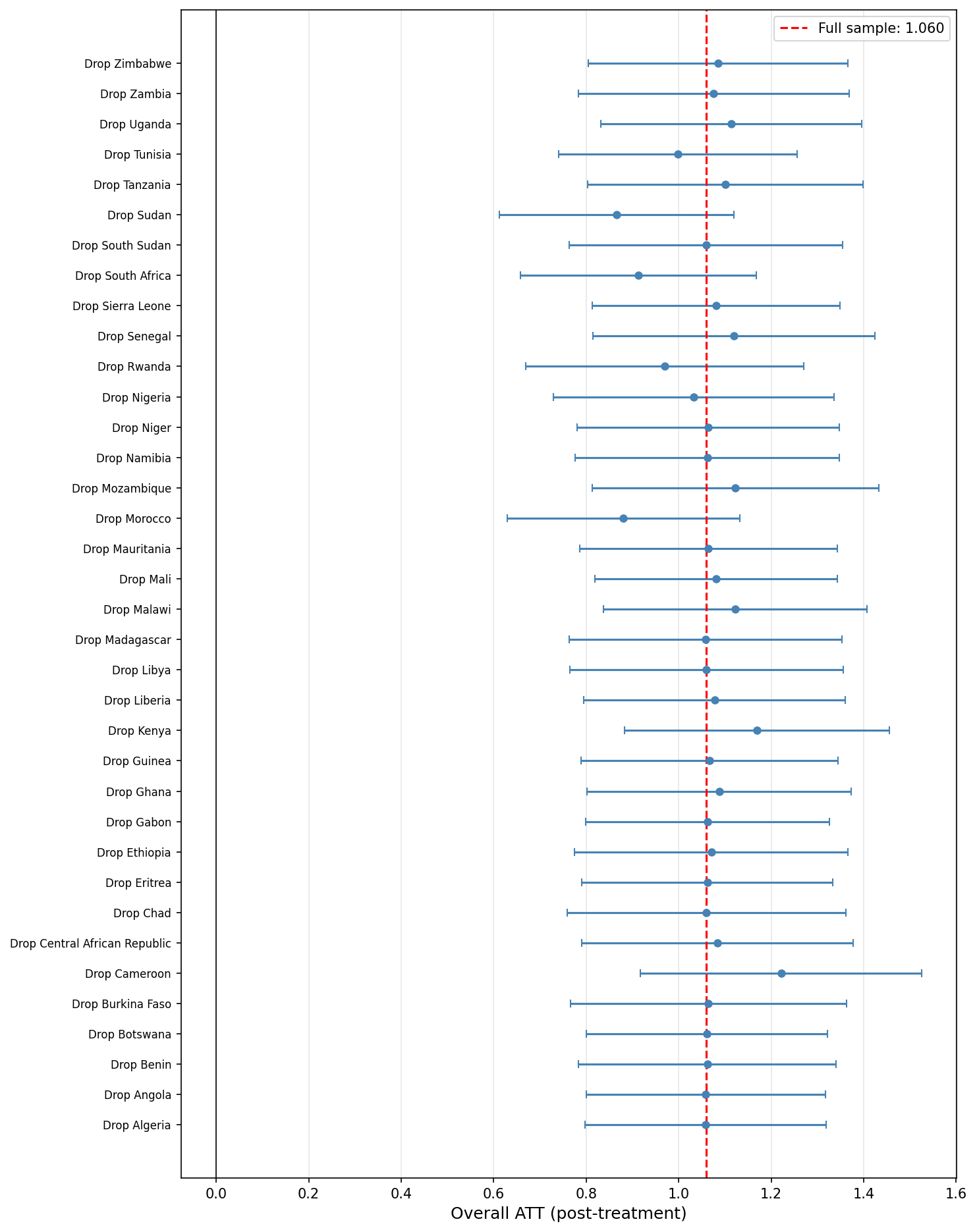}
    \caption{Leave-One-Country-Out Robustness}
    \label{fig:loco}
    \caption*{\textit{Notes.} Each point plots the ATT (protests, $50\,\text{km}$ buffer) from the imputation estimator after dropping all deals in a single country. Countries are ordered by estimated ATT. Vertical bars are 95\% confidence intervals. The horizontal dashed line marks the baseline full-sample estimate.}
\end{figure}

Figure~\ref{fig:loco} presents coefficients from a leave-one-country-out analysis. Each point represents the estimated ATT from a separate regression that drops all deals located in a given country. The narrow band of estimates indicates that the baseline result is not driven by any single country in the sample, and the lower bound of the distribution remains statistically and economically significant throughout.

\section{Sensitivity to Econometric Choices}
\label{sec:appendix_econometrics}

This appendix section reports the additional estimator comparison, the TWFE comparison, a clustering robustness check, and a standard joint pre-trend test as alternative ways of examining the main result.

\subsection{Three-Estimator Comparison}
\label{sec:appendix_three_estimator}

Table~\ref{tab:three_panel_results} presents a full three-way estimator comparison including the \citet{callaway2021difference} group-time estimator (Panel A), the imputation estimator of \citet{borusyak2024revisiting} (Panel B, the preferred specification), and the LP-DiD estimator of \citet{dube2025local} (Panel C). The imputation and LP-DiD estimates are consistent throughout.

I note that the Callaway--Sant'Anna estimates for the all-conflict outcome exhibit a marginal pre-trend ($F = 2.27$, $p = 0.05$ in the preferred country-by-year column). This likely reflects a sensitivity of the CS group-time estimator to the unbalanced panel structure introduced by the control group, which contains fewer pre-treatment periods for late-entering cohorts. The imputation estimator, which conditions directly on the counterfactual imputation step, produces clean pre-trends for both outcomes and is the preferred specification.

\input{tables/A5_baseline_3}

\subsection{TWFE vs.\ Imputation Estimator}
\label{sec:appendix_twfe}

Recent advances in the difference-in-differences literature have highlighted the potential biases of standard OLS Two-Way Fixed Effects (TWFE) estimators when treatment timing is staggered and treatment effects are heterogeneous \citep{dechaisemartin2020two,goodman2021difference}. In the main text, I rely on the \citet{borusyak2024revisiting} imputation estimator to address these issues.

Columns (1) and (2) of Table~\ref{tab:robustness_master} present the main results for the protest outcome within a $50\,\text{km}$ buffer, estimated under both the imputation and standard TWFE specifications. In the final analytical sample, the TWFE point estimate is substantially attenuated and statistically insignificant ($0.560$, SE $0.546$) while the imputation estimate remains strongly positive and tightly estimated at $1.438$ (SE $0.192$). The imputation estimate is roughly $2.6\times$ the TWFE estimate, and the TWFE estimate is no longer statistically distinguishable from zero. This pattern is consistent with the downward bias induced by negative weighting of later-treated units under staggered treatment timing. The comparison illustrates why modern heterogeneity-robust estimators are necessary in this setting and reaffirms the choice of the imputation estimator as the preferred specification.

\input{tables/A6_robustness_sweep}

\subsection{Clustering Robustness}
\label{sec:appendix_clustering}

Standard errors in the main table are clustered at the deal level using a cluster bootstrap. Because a substantial share of the sample concentrates in a small number of countries, one might worry that country-year shocks (droughts, elections, regime changes) induce within-country correlation that the deal-level clustering does not capture. Columns (4) through (6) of Table~\ref{tab:robustness_master} re-estimate the baseline specification under alternative clustering schemes: (i) deal-level (baseline); (ii) country-level; and (iii) country $\times$ year-level. Point estimates are identical across the three schemes (as expected, since only the bootstrap resampling unit changes). In the final analytical sample, the alternative clustering schemes yield slightly smaller standard errors than the deal-level baseline, and the ATT remains highly significant in all cases. The main result is therefore robust to the choice of clustering level.

\subsection{Conley Spatial HAC: Bandwidth Robustness}
\label{sec:appendix_conley_bandwidth}

The Conley spatial HAC standard errors reported alongside the cluster bootstrap in the main-text tables use a Bartlett kernel with a 200\,km bandwidth, following the \citet{harari2018conflict} convention of roughly twice the primary buffer radius. Table~\ref{tab:conley_bandwidth} reports the baseline BJS imputation ATT for protests and all-conflict events at alternative bandwidths of 100, 150, 200, 300, and 500\,km. Point estimates are fixed across columns (only the variance changes with bandwidth). For protests, the Conley SE rises modestly from $0.543$ at 100\,km to about $0.587$--$0.599$ at 200--300\,km and then falls slightly to $0.585$ at 500\,km; the estimate remains statistically significant throughout. For all-conflict events, the Conley SE rises from $1.128$ at 100\,km to $1.468$ at 500\,km, but the all-conflict estimate also remains significant across the full bandwidth range. The qualitative conclusions are therefore insensitive to the assumed radius of spatial correlation.

\input{tables/A7_conley_bandwidth}

\subsection{Standard Joint Pre-trend Tests}
\label{sec:appendix_pretrend}

The pre-trend $F$-statistic reported throughout the paper is the average squared $t$-statistic across pre-treatment event-time leads, with its joint $p$-value computed via the bootstrap. For comparability with alternative reporting conventions, Table~\ref{tab:pretrend_joint_f} reports a standard joint $F$-test of the null that all pre-treatment event-time coefficients are jointly zero, separately for each column of the main result. The qualitative conclusions are unchanged: the main imputation-DiD specifications satisfy the joint-zero pre-trend restriction at conventional levels across all four buffer radii in Table~\ref{tab:pretrend_joint_f}. The Callaway--Sant'Anna pre-trend evidence is reported separately in Table~\ref{tab:three_panel_results}, where the all-conflict CS specification remains the only column with a marginal pre-trend.

\input{tables/A8_pretrend}

\section{Spatial Sensitivity}
\label{sec:appendix_spatial}

\subsection{Buffer Sensitivity}
\label{sec:appendix_buffer_sensitivity}

A comprehensive buffer sweep across $\{25, 50, 75, 100\}\,\text{km}$ for all main results is reported in Table~\ref{tab:buffer_sweep_master} (Panel A) below. The majority of coefficients remain robust across these radii, with the clearest and most stable estimates rising monotonically for the conflict and media outcomes. The pattern therefore appears not to be an artifact of any single regional catchment choice.

\subsection{Buffer Full Radius Grid}
\label{sec:appendix_buffer_sweep}

To characterize buffer sensitivity across outcome families, I report the sweep $\{25, 50, 75, 100\}\,\text{km}$ for every outcome used in the paper in Table~\ref{tab:buffer_sweep_master}. Panel A reports the BJS imputation ATT on ACLED protests and all-conflict events. Panel C reports the Afrobarometer traditional-authority outcomes (contact and trust). The corresponding GDELT theme coverage sweep is reported in Panel B, and the electoral backlash sweep is reported in Panel D. The ACLED estimates remain positive across these radii, the GDELT theme effects remain robustly positive and rise monotonically with the catchment radius, the Afrobarometer trust response attenuates with buffer width while the contact response strengthens, and the electoral backlash patterns remain strongest for opposition support and turnout.

\input{tables/A9_buffer_sweep_master}

\subsection{Later-Treated Timing Robustness}
\label{sec:appendix_later_treated}

Table~\ref{tab:later_treated_parallel} in the main text reports a timing-based robustness check that keeps the preferred ACLED specification fixed while replacing the control group. The baseline failed-deal design compares implemented deals to failed deals that appear exogenous to local resistance. The later-treated design instead discards failed deals entirely and restricts the sample to eventually implemented projects. For each treated cohort, the comparison group is formed by deals implemented five to ten years later, and only those later cohorts' not-yet-treated observations are retained. Identification therefore comes from variation in implementation timing among successful projects rather than from implemented-versus-failed contrasts.

This is a demanding robustness check because it removes the paper's preferred failure-based counterfactual and relies only on timing variation within the universe of successful projects. At the same time, the rest of the specification is intentionally held constant: the outcome remains the cumulative ACLED count within a $50\,\text{km}$ buffer, treatment begins at negotiation, the estimator is the \citet{borusyak2024revisiting} imputation DiD, and the regression includes deal fixed effects, country $\times$ year fixed effects, and rainfall controls. In the final analytical sample, the later-treated stacked design uses 908 unique implemented deals organized into 28 cohort stacks.

\subsection{High-Spatial-Accuracy Sample Restriction}
\label{sec:appendix_spatial_accuracy_restriction}

As a stricter geolocation robustness check, I re-estimate the main ACLED specifications on the high-spatial-accuracy restriction used in the separate spatial-accuracy estimations. This restriction keeps the same outcomes, the same $50\,\text{km}$ spatial focus, the same \citet{greenstone2010agglomeration} control logic, and the same \citet{borusyak2024revisiting} imputation design, but limits the sample to deals carrying the tightest retained Land Matrix coordinate labels. The purpose is not to replace the preferred subnational-accuracy sample, but to verify that the headline protest result survives when the coordinate screen is tightened further.

\input{tables/A10_spatial_accuracy_baseline}

\begin{figure}[t]
    \centering
    \includegraphics[width=0.7\textwidth]{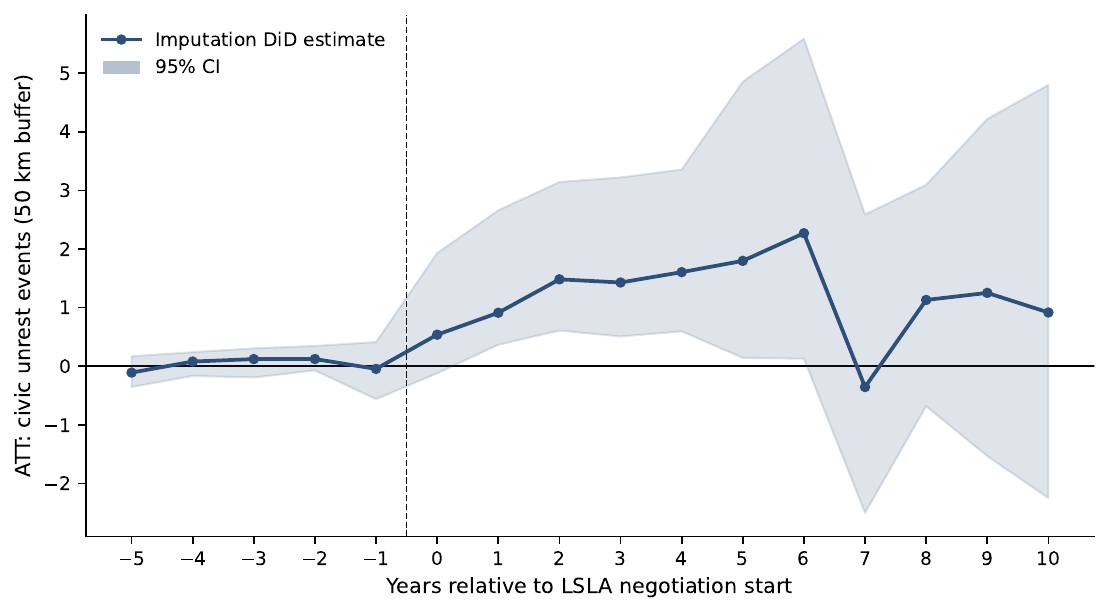}
    \caption{Dynamic ACLED Protest Effects under the High-Spatial-Accuracy Restriction}
    \label{fig:spatial_accuracy_event_study}
    \caption*{\textit{Notes.} Event-study coefficients from the \citet{borusyak2024revisiting} imputation estimator for civic protest activity within a $50\,\text{km}$ buffer under the high-spatial-accuracy restriction. The baseline comparison sample consists of implemented deals and failed or abandoned deals with no deal-specific evidence that community resistance caused the failure; the imputation uses all untreated observations, including pre-treatment observations of eventually treated units. 95\% confidence intervals are clustered at the deal level using 200 bootstrap iterations.}
\end{figure}

Table~\ref{tab:spatial_accuracy_baseline} shows that the protest and all-conflict estimates remain positive in the high-spatial-accuracy restriction, with the preferred country-by-year protest ATT equal to $1.190$ and a clean pre-trend diagnostic. Figure~\ref{fig:spatial_accuracy_event_study} complements the table by plotting the corresponding dynamic protest path, which remains flat before treatment and rises only after implementation.

\input{tables/A11_spatial_accuracy_later_treated}

Table~\ref{tab:spatial_accuracy_later_treated} then applies the later-treated timing design to the same high-spatial-accuracy subset. The civic-unrest estimate remains close to the failed-deal benchmark and the dynamic profile again shows no visible pre-treatment drift. Taken together, these tighter-coordinate results support the interpretation that the main ACLED findings are not caused by measurement error through noisier subnational geocoding.

\subsection{Early-Cohort Exclusion: ACLED Coverage Robustness}
\label{sec:appendix_early_cohort_exclusion}

ACLED's Africa panel nominally begins in 1997, but coverage is materially thinner before approximately 2002. The early years rely on a narrower set of media sources, so the absence of a recorded event in the late 1990s is more likely to reflect non-reporting than genuine calm. Because some treated cohorts in the baseline sample are assigned pre-treatment windows that overlap with this low-coverage period, an under-recorded pre-treatment mean for those cohorts could in principle inflate the imputed counterfactual and thereby bias the ATT in an indeterminate direction. 

Table~\ref{tab:early_cohort_exclusion_check} addresses this concern directly by re-estimating the six baseline specifications after dropping all implemented and failed deals with a negotiation start year before 2003. Panel A reproduces the full-sample baseline (1{,}107 treated, 284 control); Panel B restricts the sample to cohorts whose entire $-5$ to $+10$ pre/post window falls within the period of mature ACLED coverage (766 treated, 215 control).

\input{tables/A12_early_cohort_exclusion}

The restricted ATTs in Panel B remain positive and statistically significant across both outcomes and all three FE/control variants. In the preferred Country~$\times$~Year~FE plus rainfall specification, the protests ATT is $1.531$ (versus $1.484$ in the full sample) and the all-conflict ATT is $2.756$ (versus $3.080$). The civic-unrest estimate is therefore slightly larger after dropping early cohorts, and the all-conflict estimate is modestly smaller; both remain close to the full-sample baseline. Pre-trend $F$ statistics under the preferred Country~$\times$~Year~FE specifications are clean in both panels ($p \geq 0.42$). The headline conclusion is therefore robust to the ACLED early-coverage concern.

\subsection{Treated-Buffer Overlap and SUTVA}
\label{sec:appendix_spillovers}

A potential concern with the $50\,\text{km}$ baseline buffer is that treated catchments overlap when deals are clustered in the same region, raising a SUTVA concern. Table~\ref{tab:buffer_overlap} reports the share of treated deals whose buffer overlaps with at least one other treated deal, together with the average number of overlapping neighbors among deals with any overlap, for buffers of $25\,\text{km}$, $50\,\text{km}$, and $75\,\text{km}$. At the $50\,\text{km}$ baseline, 92.1\% of treated deals have at least one neighbor within the overlap threshold ($2 \times 50 = 100\,\text{km}$), reflecting the geographic clustering of agricultural investment zones across Africa. 

\input{tables/A13_buffer_overlap}

While this overlap rate is high, the deal-level fixed effects in the imputation estimator absorb any time-invariant location-specific conflict level, so the identifying variation comes from within-deal temporal changes rather than cross-sectional comparisons across overlapping buffers. The consistency of the ATT across the $25\,\text{km}$ buffer (where overlap is lower) and the $75\,\text{km}$ buffer (where overlap is higher) in Table~\ref{tab:sensitivity_specs} provides further reassurance that spatial spillovers across treated deals do not materially affect the estimated effect.

\input{tables/A14_drop_clustered}

As a direct SUTVA robustness check, Table~\ref{tab:drop_clustered} re-estimates the baseline BJS imputation ATT after dropping treated deals with many nearby treated neighbors. The top-decile cut (threshold $>20$ neighbors, 91 deals dropped) slightly decreases the point estimate to $1.404$. Dropping any deal with more than ten overlapping neighbors yields $1.308$, still close to baseline, while the most aggressive cut, dropping any deal with more than five neighbors, raises the estimate to $1.684$ on the small retained sample. The estimate therefore remains stable around the baseline across progressively more aggressive trimming rules, with the most aggressive cut showing a modestly larger effect on the smaller retained sample. The evidence points against overlap-driven inflation of the main estimate.

\section{Supplementary Conflict and Mechanism Results}
\label{sec:appendix_mechanisms}

\subsection{Armed Conflict Outcomes}
\label{sec:appendix_armed_conflict}

The main text emphasizes civic mobilization as the paper's cleanest and most theoretically direct response to LSLA implementation. To clarify the composition of the broader security response, this appendix subsection reports a supplementary armed-conflict decomposition using ACLED outcome categories. I define the armed-conflict composite as
\[
    n_{\text{armed conflict}} = n_{\text{battles}} + n_{\text{explosions}} + n_{\text{vac}},
\]
where the component terms correspond respectively to Battles, Explosions/Remote violence, and Violence against civilians. Strategic developments and fatalities are reported separately as diagnostics.

\input{tables/A15_armed_conflict_outcomes}

\begin{figure}[t]
    \centering
    \includegraphics[width=0.8\textwidth]{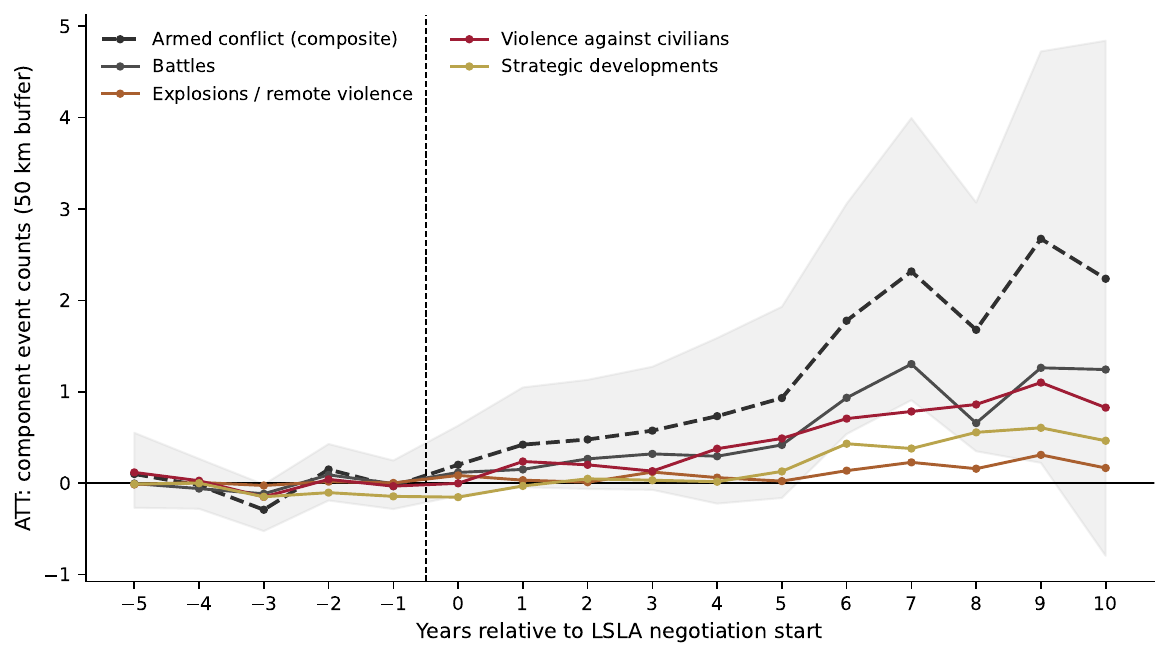}
    \caption{Supplementary Armed-Conflict Outcome Decomposition}
    \label{fig:armed_conflict_appendix}
    \caption*{\textit{Notes.} Dynamic event-study estimates from the imputation estimator for the armed-conflict composite and its constituent ACLED categories within a $50\,\text{km}$ buffer. The composite sums Battles, Explosions/Remote violence, and Violence against civilians. Strategic developments are shown as an additional diagnostic. All specifications use the failed-deal control group with community-resistance cases excluded. 95\% confidence intervals are clustered at the deal level.}
\end{figure}

Table~\ref{tab:armed_conflict_outcomes} reports the aggregate post-treatment ATT, standard error, confidence interval, pre-trend $F$-statistic, and pre-treatment mean for the protest benchmark, the armed-conflict composite, and the individual component outcomes. The protest row is included only as a benchmark for scale, showing its large effect size relative to its pre-treatment mean; the remaining rows are intended as an outcome decomposition of the broader security response. Figure~\ref{fig:armed_conflict_appendix} complements the table by plotting the dynamic path of the armed-conflict composite together with its constituent categories.

\subsection{Heterogeneity: Ownership Robustness}
\label{sec:appendix_heterogeneity}

Figure~\ref{fig:heterogeneity_asset} in the main text reports the failed-control event-study plots for investor origin and crop type, while Table~\ref{tab:ownership_later_treated_main} reports the strict mutually exclusive ownership comparison under the later-treated design.

\input{tables/A16_heterogeneity_full}

Table~\ref{tab:heterogeneity_full} reports the corresponding ownership robustness estimates for broader non-exclusive prior-owner definitions under the same later-treated design. The goal is to show that the ownership pattern is not unique to one especially strict coding rule: broader definitions that include community, indigenous, or smallholder-linked prior claims also yield sizeable post-treatment increases in civic unrest, while state-linked land remains positively associated with protest under the same timing-based comparison.

%% file: tables/A1_control_sample_waterfall.tex
\begin{table}[tbp]
\centering
\caption{Analytical Sample Construction and Failed-Control Screening}
\label{tab:control_sample_waterfall}
\begin{threeparttable}
\setlength{\tabcolsep}{6pt}
\begin{tabularx}{\textwidth}{>{\raggedright\arraybackslash}X rrr}
\toprule
Stage & Deals & Dropped & Share (\%) \\
\midrule
\multicolumn{4}{l}{\textbf{Panel A: Analytical sample construction}} \\
\addlinespace[3pt]
Global Land Matrix deals & 7,671 & --- & 100.0 \\
Africa-targeted deals & 2,349 & 5,322 & 30.6 \\
Deals in the 38 study countries & 2,109 & 240 & 89.8 \\
Deals in the study countries with valid coordinates & 1,840 & 269 & 87.2 \\
Deals entering the implemented-vs-failed comparison sample & 1,458 & 382 & 79.2 \\
Implemented deals retained as treated units & 1,107 & --- & 75.9 \\
Failed/abandoned candidate controls & 351 & --- & 24.1 \\
Failed controls retained after endogenous-failure screening & 284 & 67 & 80.9 \\
Final analytical sample & 1,391 & --- & 95.4 \\
\midrule
\multicolumn{4}{l}{\textbf{Panel B: Failed-control screening}} \\
\addlinespace[3pt]
Failed/abandoned candidate controls & 351 & --- & 100.0 \\
After excluding documented community resistance or local land conflict & 293 & 58 & 83.5 \\
After supplementary source-first review of previously unresolved failures & 284 & 9 & 80.9 \\
Confirmed-exogenous controls used only in A.2 & 83 & 201 & 23.6 \\
\bottomrule
\end{tabularx}
\begin{tablenotes}[flushleft]
\footnotesize
\item \textit{Notes.} Panel~A traces the current live comparison sample from the raw Land Matrix universe to the geocoded ACLED deal panel. The geocoding row retains study-country deals with at least one non-missing latitude-longitude pair in the processed LSLA file. Rows that split implemented and failed deals report branches of the comparison sample rather than additional sequential drops. Panel~B reports the sequential screening of the failed or abandoned candidate controls in the precision sample. Community-resistance exclusions are identified from free-text negotiation and implementation comments, the structured \textit{Presence of land conflicts} field, the structured \textit{Community reaction} field, and a supplementary source-first review of previously unresolved failed deals. The final row reports the subset of failed controls that also satisfy the stricter confirmed-exogenous rule used only in Table~\ref{tab:confirmed_exogenous_controls}.
\end{tablenotes}
\end{threeparttable}
\end{table}

%% file: tables/A2_confirmed_exogenous_controls.tex
\begin{table}[tbp]
\centering
\caption{Effect of Large-Scale Land Acquisitions on Local Conflict under Confirmed Exogenous Controls}
\label{tab:confirmed_exogenous_controls}
\begin{threeparttable}
\setlength{\tabcolsep}{10pt}
\begin{tabular}{lD{.}{.}{3}D{.}{.}{3}D{.}{.}{3}D{.}{.}{3}D{.}{.}{3}D{.}{.}{3}}
\toprule
\textit{Dependent Variable:} & \multicolumn{3}{c}{Protests} & \multicolumn{3}{c}{All conflict events} \\
\cmidrule(lr){2-4} \cmidrule(lr){5-7}
 & \multicolumn{1}{c}{(1)} & \multicolumn{1}{c}{(2)} & \multicolumn{1}{c}{(3)} & \multicolumn{1}{c}{(4)} & \multicolumn{1}{c}{(5)} & \multicolumn{1}{c}{(6)} \\
\midrule
ATT (post-treatment) & 0.405^{*} & 1.258^{***} & 1.276^{***} & 2.242^{***} & 3.028^{***} & 3.039^{***} \\
 & (0.224) & (0.215) & (0.219) & (0.343) & (0.359) & (0.369) \\
 & [0.652] & [0.594] & [0.630] & [1.535] & [1.339] & [1.387] \\
\addlinespace[4pt]
Pre-trend $F$ & 0.77 & 1.04 & 0.95 & 1.02 & 0.75 & 0.84 \\
 & [0.57] & [0.39] & [0.45] & [0.41] & [0.59] & [0.52] \\
\midrule
Pre-treatment mean & \multicolumn{1}{c}{0.94} & \multicolumn{1}{c}{0.94} & \multicolumn{1}{c}{0.94} & \multicolumn{1}{c}{3.90} & \multicolumn{1}{c}{3.90} & \multicolumn{1}{c}{3.90} \\
$N$ (treated) & \multicolumn{1}{c}{1,107} & \multicolumn{1}{c}{1,107} & \multicolumn{1}{c}{1,107} & \multicolumn{1}{c}{1,107} & \multicolumn{1}{c}{1,107} & \multicolumn{1}{c}{1,107} \\
$N$ (control) & \multicolumn{1}{c}{83} & \multicolumn{1}{c}{83} & \multicolumn{1}{c}{83} & \multicolumn{1}{c}{83} & \multicolumn{1}{c}{83} & \multicolumn{1}{c}{83} \\
\midrule
Deal FE & \multicolumn{1}{c}{\checkmark} & \multicolumn{1}{c}{\checkmark} & \multicolumn{1}{c}{\checkmark} & \multicolumn{1}{c}{\checkmark} & \multicolumn{1}{c}{\checkmark} & \multicolumn{1}{c}{\checkmark} \\
Year FE & \multicolumn{1}{c}{\checkmark} &  &  & \multicolumn{1}{c}{\checkmark} &  &  \\
Country $\times$ Year FE &  & \multicolumn{1}{c}{\checkmark} & \multicolumn{1}{c}{\checkmark} &  & \multicolumn{1}{c}{\checkmark} & \multicolumn{1}{c}{\checkmark} \\
Rainfall control &  &  & \multicolumn{1}{c}{\checkmark} &  &  & \multicolumn{1}{c}{\checkmark} \\
\bottomrule
\end{tabular}
\begin{tablenotes}[para,flushleft]
\footnotesize
\item \textit{Notes.} Each column is a separate event study using the \citet{borusyak2024revisiting} imputation estimator. The unit of observation is an LSLA deal, with outcomes aggregated within a 50\,km buffer. The baseline comparison sample consists of implemented deals ($N=1,107$) and failed deals in the confirmed-exogenous subset ($N=83$); the imputation uses all untreated observations, including pre-treatment observations of eventually treated units. The protests outcome counts ACLED ``Protests'' and ``Riots'' events within the buffer. The all-conflict outcome counts all ACLED event types (Battles, Protests, Riots, Violence against civilians, Explosions/Remote violence, and Strategic developments). Bootstrapped standard errors (200 iterations) clustered by deal are reported in parentheses, and \citet{conley1999gmm} spatial HAC standard errors are reported in brackets (Bartlett kernel, 200\,km bandwidth). ATT is the average post-treatment effect ($k \geq 0$). Pre-trend $F$ is the average squared $t$-statistic for $k < 0$; joint $p$-values are reported in brackets. \\
\item $^{*}p<0.10$, $^{**}p<0.05$, $^{***}p<0.01$
\end{tablenotes}
\end{threeparttable}
\end{table}

%% file: tables/A3_failure_prediction.tex
\begin{table}[tbp]
\centering
\caption{Logit Predictors of Deal Abandonment}
\label{tab:failure_prediction}
\begin{threeparttable}
\setlength{\tabcolsep}{10pt}
\begin{tabular}{lc}
\toprule
\textit{Dependent Variable:} & Pr(Deal Failed = 1) \\
 & (1) \\
\midrule
\multicolumn{2}{l}{\textbf{Panel A: Pre-Treatment Local Conflict}} \\
\quad Protests \& Riots & -0.002 \\
 & (0.007) \\
\quad All Conflict Events & $-0.004^{*}$ \\
 & (0.002) \\
\addlinespace
\midrule
\multicolumn{2}{l}{\textbf{Panel B: Geographic Fundamentals}} \\
\quad Log distance to capital (km) & 0.006 \\
 & (0.011) \\
\midrule
 Pseudo-$R^2$ & 0.138 \\
 $N$ (treated) & 1,057 \\
 $N$ (control)  & 280 \\
\midrule
 Country FE & \checkmark \\
\bottomrule
\end{tabular}
\begin{tablenotes}[para,flushleft]
\footnotesize
\item \textit{Notes.} Average marginal effects from a standard logit regression of an indicator for failed deal status on pre-treatment covariates, evaluated at the sample mean. Standard errors (in parentheses) are from 200 bootstrap iterations. The logit is estimated on the subset of countries with within-country variation in deal failure status, which is the estimable sample for country fixed effects. Pre-treatment conflict averages are computed from ACLED events within a 50km buffer over the five years prior to the deal's negotiation year. Country fixed effects absorb differences in Land Matrix documentation intensity. The refreshed fixed-effects logit does not show a clean pattern in which higher pre-treatment conflict systematically predicts failure. The protest coefficient is weakly positive, while the broader all-events coefficient is weakly negative. This mixed pattern is difficult to reconcile with a simple story in which the control group is selected on latent local resistance. \\
\item $^{*}p<0.10$, $^{**}p<0.05$, $^{***}p<0.01$
\end{tablenotes}
\end{threeparttable}
\end{table}

%% file: tables/A4_composition_balanced.tex
\begin{table}[tbp]
\centering
\caption{Composition-Balanced ATT under Investor-Origin Adjustments}
\label{tab:composition_balanced}
\begin{threeparttable}
\setlength{\tabcolsep}{4pt}
\begin{tabular}{lccccc}
\toprule
 & \multicolumn{1}{c}{Baseline} & \multicolumn{2}{c}{Subsample} & \multicolumn{1}{c}{IPW} & \multicolumn{1}{c}{Control Type} \\
\cmidrule(lr){2-2} \cmidrule(lr){3-4} \cmidrule(lr){5-5} \cmidrule(lr){6-6}
 & (1) & (2) & (3) & (4) & (5) \\
 & All & Domestic & Transnational & Reweighted & Drop Admin \\
\midrule
 ATT & $1.438^{***}$ & $1.786^{***}$ & $1.396^{***}$ & $1.546^{***}$ & $1.448^{***}$ \\
 & (0.192) & (0.198) & (0.244) & (0.191) & (0.206) \\
\midrule
 $N$ (treated) & 1,107 & 432 & 675 & 1,107 & 1,107 \\
 $N$ (control) & 284 & 45 & 239 & 284 & 276 \\
 \midrule
Deal FE & \multicolumn{1}{c}{\checkmark} & \multicolumn{1}{c}{\checkmark} & \multicolumn{1}{c}{\checkmark} & \multicolumn{1}{c}{\checkmark} & \multicolumn{1}{c}{\checkmark} \\
Country $\times$ Year FE & \multicolumn{1}{c}{\checkmark} & \multicolumn{1}{c}{\checkmark} & \multicolumn{1}{c}{\checkmark} & \multicolumn{1}{c}{\checkmark} & \multicolumn{1}{c}{\checkmark} \\
\bottomrule
\end{tabular}
\begin{tablenotes}[para,flushleft]
\footnotesize
\item \textit{Notes.} Pooled post-treatment ATT from the \citet{borusyak2024revisiting} imputation estimator on the 50\,km buffer panel. Outcome: annual count of civic protests and riots. Column (1) reproduces the main specification (Country $\times$ Year FE). Columns (2)--(3) restrict \emph{both} treated and control groups to the same investor-origin category before estimation, eliminating compositional imbalance within each stratum. Column (4) reweights control deals so their investor-origin distribution matches the treated group's domestic/transnational composition; treated deals receive weight one. Column (5) decomposes the control group by failure reason, dropping controls whose documented failure involves administrative or government action (permit/license revocations, ministerial denials, court orders, moratoria, expropriation), keeping only investor-side and contract-expiry failures. Cluster bootstrap SE (200 iterations, clustered at deal) in parentheses. \\
\item $^{*}p<0.10$, $^{**}p<0.05$, $^{***}p<0.01$
\end{tablenotes}
\end{threeparttable}
\end{table}

%% file: tables/A5_baseline_3.tex
\begin{table}[tbp]
\centering
\caption{Effect of Large-Scale Land Acquisitions on Local Conflict (Three Estimators)}
\label{tab:three_panel_results}
\begin{threeparttable}
\setlength{\tabcolsep}{10pt}
\begin{tabular}{lD{.}{.}{3}D{.}{.}{3}D{.}{.}{3}D{.}{.}{3}D{.}{.}{3}D{.}{.}{3}}
\toprule
\textit{Dependent Variable:} & \multicolumn{3}{c}{Protests} & \multicolumn{3}{c}{All conflict events} \\
\cmidrule(lr){2-4} \cmidrule(lr){5-7}
 & \multicolumn{1}{c}{(1)} & \multicolumn{1}{c}{(2)} & \multicolumn{1}{c}{(3)} & \multicolumn{1}{c}{(4)} & \multicolumn{1}{c}{(5)} & \multicolumn{1}{c}{(6)} \\
\midrule
\multicolumn{7}{l}{\textbf{Panel A: Callaway–Sant'Anna estimator}} \\
\addlinespace[3pt]
ATT (post-treatment) & 1.009^{**} & 1.009^{**} & 1.184^{**} & 3.497^{***} & 3.497^{***} & 3.815^{***} \\
 & (0.470) & (0.420) & (0.473) & (0.655) & (0.813) & (0.837) \\
\addlinespace[4pt]
Pre-trend $F$ & 0.69 & 0.97 & 0.86 & 1.89 & 1.90 & 2.27 \\
 & [0.63] & [0.43] & [0.51] & [0.09] & [0.09] & [0.05] \\
\midrule
\multicolumn{7}{l}{\textbf{Panel B: Imputation DiD estimator}} \\
\addlinespace[3pt]
ATT (post-treatment) & 1.106^{***} & 1.438^{***} & 1.484^{***} & 2.904^{***} & 2.915^{***} & 3.080^{***} \\
 & (0.207) & (0.192) & (0.200) & (0.334) & (0.351) & (0.372) \\
\addlinespace[4pt]
Pre-trend $F$ & 1.89 & 1.33 & 1.38 & 0.64 & 0.66 & 0.66 \\
 & [0.09] & [0.25] & [0.23] & [0.67] & [0.65] & [0.65] \\
\midrule
\multicolumn{7}{l}{\textbf{Panel C: LP-DiD estimator}} \\
\addlinespace[3pt]
ATT (post-treatment) & 1.137^{**} & 1.111^{***} & 1.180^{***} & 3.287^{***} & 2.959^{***} & 3.183^{***} \\
 & (0.441) & (0.420) & (0.434) & (0.672) & (0.693) & (0.719) \\
\addlinespace[4pt]
Pre-trend $F$ & 0.75 & 0.74 & 0.67 & 0.51 & 0.51 & 0.55 \\
 & [0.56] & [0.57] & [0.62] & [0.73] & [0.72] & [0.70] \\
\midrule
Pre-treatment mean & \multicolumn{1}{c}{0.94} & \multicolumn{1}{c}{0.94} & \multicolumn{1}{c}{0.94} & \multicolumn{1}{c}{3.90} & \multicolumn{1}{c}{3.90} & \multicolumn{1}{c}{3.90} \\
$N$ (treated) & \multicolumn{1}{c}{1,107} & \multicolumn{1}{c}{1,107} & \multicolumn{1}{c}{1,107} & \multicolumn{1}{c}{1,107} & \multicolumn{1}{c}{1,107} & \multicolumn{1}{c}{1,107} \\
$N$ (control) & \multicolumn{1}{c}{284} & \multicolumn{1}{c}{284} & \multicolumn{1}{c}{284} & \multicolumn{1}{c}{284} & \multicolumn{1}{c}{284} & \multicolumn{1}{c}{284} \\
\midrule
Deal FE & \multicolumn{1}{c}{\checkmark} & \multicolumn{1}{c}{\checkmark} & \multicolumn{1}{c}{\checkmark} & \multicolumn{1}{c}{\checkmark} & \multicolumn{1}{c}{\checkmark} & \multicolumn{1}{c}{\checkmark} \\
Year FE & \multicolumn{1}{c}{\checkmark} &  &  & \multicolumn{1}{c}{\checkmark} &  &  \\
Country ($\times$ Year) FE\tnote{a} &  & \multicolumn{1}{c}{\checkmark} & \multicolumn{1}{c}{\checkmark} &  & \multicolumn{1}{c}{\checkmark} & \multicolumn{1}{c}{\checkmark} \\
Rainfall control &  &  & \multicolumn{1}{c}{\checkmark} &  &  & \multicolumn{1}{c}{\checkmark} \\
\bottomrule
\end{tabular}
\begin{tablenotes}[para,flushleft]
\footnotesize
\item \textit{Notes.} Each column is a separate event study. The unit of observation is an LSLA deal, with outcomes aggregated within a 50\,km buffer. The baseline comparison sample consists of implemented deals ($N=1,107$) and failed deals ($N=284$). Panel A uses the group-time estimator of \citet{callaway2021difference} with never-treated controls, outcome-regression adjustment, and 200 multiplier-bootstrap iterations. Panel B uses the imputation estimator of \citet{borusyak2024revisiting}; its imputation uses all untreated observations, including pre-treatment observations of eventually treated units. Panel C uses the LP-DiD estimator of \citet{dube2025local}. Standard errors in parentheses are clustered at the deal level. ATT is the average post-treatment effect ($k \geq 0$). Pre-trend $F$ is the average squared $t$-statistic for $k < 0$; joint $p$-values are reported in brackets. \\
\item $^{*}p<0.10$, $^{**}p<0.05$, $^{***}p<0.01$ \\
\item[a] Panel A: country dummies enter the CS outcome-regression step via \texttt{xformla}. Panel B: country $\times$ year fixed effects absorbed in the imputation step. Panel C: country dummies included as covariates in the LP-DiD first-differenced regression.
\end{tablenotes}
\end{threeparttable}
\end{table}

%% file: tables/A6_robustness_sweep.tex
\begin{table}[tbp]
\centering
\caption{Robustness to Alternative Specifications and Subsamples}
\label{tab:robustness_master}
\begin{threeparttable}
\setlength{\tabcolsep}{6pt}
\begin{tabular}{lcccccc}
\toprule
 & \multicolumn{2}{c}{Estimator} & \multicolumn{1}{c}{Sample} & \multicolumn{3}{c}{Clustering} \\
\cmidrule(lr){2-3} \cmidrule(lr){4-4} \cmidrule(lr){5-7}
 & (1) & (2) & (3) & (4) & (5) & (6) \\
 & Baseline & TWFE & Tightened & Deal & Country & C $\times$ Y \\
\midrule
  ATT & $1.438^{***}$ & $0.560$ & $1.351^{***}$ & $1.438^{***}$ & $1.438^{***}$ & $1.438^{***}$ \\
  & (0.192) & (0.546) & (0.186) & (0.192) & (0.219) & (0.189) \\
\midrule
  $N$ (treated) & 1,107 & 1,107 & 1,107 & 1,107 & 1,107 & 1,107 \\
  $N$ (control) & 284 & 284 & 266 & 284 & 284 & 284 \\
  \midrule
  Deal FE & \checkmark & \checkmark & \checkmark & \checkmark & \checkmark & \checkmark \\
  Country $\times$ Year FE & \checkmark & \checkmark & \checkmark & \checkmark & \checkmark & \checkmark \\
\bottomrule
\end{tabular}
\begin{tablenotes}[para,flushleft]
\footnotesize
\item \textit{Notes.} Pooled post-treatment ATT. Outcome: annual count of civic protests and riots ($50\,\text{km}$ buffer). All specifications use country $\times$ year FE and deal FE. Columns (1)--(2) compare the baseline \citet{borusyak2024revisiting} imputation estimator (200 cluster bootstrap iterations) to a standard TWFE regression (which is susceptible to negative weighting under staggered treatment timing \citep{dechaisemartin2020two,goodman2021difference}). Column (3) restricts the control group by additionally dropping any deal with opposition, resistance, or protest keywords in the status text. Columns (4)--(6) vary the cluster bootstrap level (200 iterations). \\
\item $^{*}p<0.10$, $^{**}p<0.05$, $^{***}p<0.01$
\end{tablenotes}
\end{threeparttable}
\end{table}

%% file: tables/A7_conley_bandwidth.tex
\begin{table}[tbp]
\centering
\caption{Conley Spatial HAC Bandwidth Robustness}
\label{tab:conley_bandwidth}
\setlength{\tabcolsep}{10pt}
\begin{threeparttable}
\begin{tabular}{lccccc}
\toprule
\textit{Bandwidth:} & 100\,km & 150\,km & 200\,km & 300\,km & 500\,km \\
 & (1) & (2) & (3) & (4) & (5) \\
\midrule
\multicolumn{6}{l}{\textbf{Panel A: Protests}} \\
\quad ATT & $1.444^{***}$ & $1.444^{**}$ & $1.444^{**}$ & $1.444^{**}$ & $1.444^{**}$ \\
\quad Conley SE & $[0.543]$ & $[0.572]$ & $[0.587]$ & $[0.599]$ & $[0.585]$ \\
\addlinespace
\quad Pre-trend $F$-stat & $1.33$ & $1.33$ & $1.33$ & $1.33$ & $1.33$ \\
\quad Pre-trend $p$-value & $0.250$ & $0.250$ & $0.250$ & $0.250$ & $0.250$ \\
\midrule
\multicolumn{6}{l}{\textbf{Panel B: All Conflict Events}} \\
\quad ATT & $2.944^{***}$ & $2.944^{**}$ & $2.944^{**}$ & $2.944^{**}$ & $2.944^{**}$ \\
\quad Conley SE & $[1.128]$ & $[1.224]$ & $[1.297]$ & $[1.378]$ & $[1.468]$ \\
\addlinespace
\quad Pre-trend $F$-stat & $0.66$ & $0.66$ & $0.66$ & $0.66$ & $0.66$ \\
\quad Pre-trend $p$-value & $0.651$ & $0.651$ & $0.651$ & $0.651$ & $0.651$ \\
\bottomrule
\end{tabular}
\begin{tablenotes}[para,flushleft]
\footnotesize
\item \textit{Notes.} Pooled post-treatment ATT from the \citet{borusyak2024revisiting} imputation estimator on the 50\,km buffer panel with country $\times$ year FE. The point estimates do not depend on the bandwidth; columns vary only the Bartlett kernel bandwidth used for the \citet{conley1999gmm} spatial HAC standard errors and the corresponding joint pre-trend tests. The main text reports the 200\,km bandwidth (following the \citet{harari2018conflict} convention of roughly twice the main buffer radius). \\
\item $^{*}p<0.10$, $^{**}p<0.05$, $^{***}p<0.01$
\end{tablenotes}
\end{threeparttable}
\end{table}

%% file: tables/A8_pretrend.tex
\begin{table}[tbp]
\centering
\caption{Pre-Trend Joint Tests Across Submission Specifications}
\label{tab:pretrend_joint_f}
\setlength{\tabcolsep}{16pt}
\begin{threeparttable}
\begin{tabular}{lcccc}
\toprule
 & \multicolumn{4}{c}{Spatial Buffer Radius} \\
\cmidrule(lr){2-5}
Statistic & 25 km & \textbf{50 km} & 75 km & 100 km \\
 & (1) & \textbf{(2)} & (3) & (4) \\
\midrule
\multicolumn{5}{l}{\textbf{Protests}} \\
\quad Avg.\ $t^2$ & 1.43 & \textbf{1.33} & 1.92 & 2.13 \\
\quad $\chi^2$ stat (5 df) & 7.17 & \textbf{6.63} & 9.59 & 10.64 \\
\quad $p$-value & 0.208 & \textbf{0.250} & 0.088 & 0.059 \\
\quad Reject at 5\%? & No & \textbf{No} & No & No \\
\addlinespace
\multicolumn{5}{l}{\textbf{All Conflict Events}} \\
\quad Avg.\ $t^2$ & 0.80 & \textbf{0.66} & 0.38 & 1.01 \\
\quad $\chi^2$ stat (5 df) & 3.99 & \textbf{3.32} & 1.91 & 5.05 \\
\quad $p$-value & 0.551 & \textbf{0.651} & 0.861 & 0.410 \\
\quad Reject at 5\%? & No & \textbf{No} & No & No \\
\bottomrule
\end{tabular}
\begin{tablenotes}[para,flushleft]
\footnotesize
\item \textit{Notes.} Joint Wald test of the null that all pre-treatment event-study coefficients equal zero. The test statistic is $k \times \overline{t^2}$, where $\overline{t^2}$ is the average squared $t$-statistic across the $k=5$ pre-treatment leads. All specifications use the \citet{borusyak2024revisiting} imputation estimator with deal fixed effects and Country $\times$ Year FE. The baseline comparison sample consists of implemented deals and failed deals with no deal-specific evidence that community resistance caused the failure; the imputation uses all untreated observations, including pre-treatment observations of eventually treated units. Standard errors come from 200 bootstrap iterations clustered at the deal level. The \textbf{50 km} column is the main-text baseline.
\end{tablenotes}
\end{threeparttable}
\end{table}

%% file: tables/A9_buffer_sweep_master.tex
\begin{table}[tbp]
\centering
\caption{Buffer Sensitivity Sweep Across All Main Outcomes}
\label{tab:buffer_sweep_master}
\setlength{\tabcolsep}{8pt}
\begin{threeparttable}
\begin{tabular}{lcccc}
\toprule
 & \multicolumn{4}{c}{Spatial Buffer Radius} \\
\cmidrule(lr){2-5}
Dependent Variable & 25 km & \textbf{50 km} & 75 km & 100 km \\
 & (1) & \textbf{(2)} & (3) & (4) \\
\midrule
\multicolumn{5}{l}{\textbf{Panel A: Conflict (ACLED)}} \\
\quad Protests \& Riots & $0.600^{***}$ & $\mathbf{1.438^{***}}$ & $1.954^{***}$ & $2.300^{***}$ \\
 & (0.116) & \textbf{(0.192)} & (0.237) & (0.331) \\
\quad All Conflict Events & $0.996^{***}$ & $\mathbf{2.915^{***}}$ & $4.538^{***}$ & $5.537^{***}$ \\
 & (0.178) & \textbf{(0.351)} & (0.706) & (1.294) \\
\addlinespace
\multicolumn{5}{l}{\textbf{Panel B: Media Discourse (GDELT GKG)}} \\
\quad Property Rights & $0.209^{***}$ & $\mathbf{0.472^{***}}$ & $0.633^{***}$ & $0.885^{***}$ \\
 & (0.026) & \textbf{(0.036)} & (0.042) & (0.050) \\
\quad Corruption & $0.667^{***}$ & $\mathbf{1.074^{***}}$ & $1.384^{***}$ & $1.733^{***}$ \\
 & (0.040) & \textbf{(0.047)} & (0.051) & (0.057) \\
\quad Agriculture & $0.814^{***}$ & $\mathbf{1.348^{***}}$ & $1.600^{***}$ & $1.887^{***}$ \\
 & (0.039) & \textbf{(0.044)} & (0.049) & (0.053) \\
\addlinespace
\multicolumn{5}{l}{\textbf{Panel C: Traditional Authority (Afrobarometer)}} \\
\quad Contact trad.\ leader & $-0.084^{**}$ & $\mathbf{-0.083^{***}}$ & $-0.096^{***}$ & $-0.103^{***}$ \\
 & (0.033) & \textbf{(0.027)} & (0.020) & (0.034) \\
\quad Trust trad.\ leaders & $-0.063^{**}$ & $\mathbf{-0.049^{***}}$ & $-0.045^{***}$ & $-0.033$ \\
 & (0.027) & \textbf{(0.018)} & (0.016) & (0.029) \\
\addlinespace
\multicolumn{5}{l}{\textbf{Panel D: Electoral Backlash (CLEA)}} \\
\quad Incumbent Vote Share & $0.004$ & $\mathbf{0.008}$ & $0.025^{***}$ & $-0.027^{***}$ \\
 & (0.013) & \textbf{(0.010)} & (0.009) & (0.009) \\
\quad Opposition Vote Share & $0.115^{***}$ & $\mathbf{0.100^{***}}$ & $0.062^{***}$ & $0.064^{***}$ \\
 & (0.013) & \textbf{(0.009)} & (0.008) & (0.008) \\
\quad Voter Turnout & $0.094^{***}$ & $\mathbf{0.088^{***}}$ & $0.083^{***}$ & $0.032^{***}$ \\
 & (0.011) & \textbf{(0.009)} & (0.008) & (0.005) \\
\bottomrule
\end{tabular}
\begin{tablenotes}[para,flushleft]
\footnotesize
\item \textit{Notes.} Pooled post-treatment ATT estimates from the \citet{borusyak2024revisiting} imputation estimator across four spatial buffer radii. The \textbf{50 km} column represents the main-text baseline. Panel A reports ACLED civic unrest and all conflict events, with deal fixed effects and Country $\times$ Year FE. Panel B reports GDELT theme coverage using the inverse hyperbolic sine of annual article counts, with deal fixed effects, Country $\times$ Year FE, and rainfall controls. Panel C reports Afrobarometer traditional-authority outcomes, with EA fixed effects and Country $\times$ Round FE. Panel D reports CLEA electoral outcomes, with constituency fixed effects and Country $\times$ election-year FE. Standard errors in parentheses come from 200 bootstrap iterations clustered at the relevant geographic unit. \\
\item $^{*}p<0.10$, $^{**}p<0.05$, $^{***}p<0.01$
\end{tablenotes}
\end{threeparttable}
\end{table}

%% file: tables/A10_spatial_accuracy_baseline.tex
\begin{table}[tbp]
\centering
\caption{Effect of Large-Scale Land Acquisitions on Local Conflict}
\label{tab:spatial_accuracy_baseline}
\begin{threeparttable}
\setlength{\tabcolsep}{10pt}
\begin{tabular}{lD{.}{.}{3}D{.}{.}{3}D{.}{.}{3}D{.}{.}{3}D{.}{.}{3}D{.}{.}{3}}
\toprule
\textit{Dependent Variable:} & \multicolumn{3}{c}{Protests} & \multicolumn{3}{c}{All conflict events} \\
\cmidrule(lr){2-4} \cmidrule(lr){5-7}
 & \multicolumn{1}{c}{(1)} & \multicolumn{1}{c}{(2)} & \multicolumn{1}{c}{(3)} & \multicolumn{1}{c}{(4)} & \multicolumn{1}{c}{(5)} & \multicolumn{1}{c}{(6)} \\
\midrule
ATT (post-treatment) & -0.148 & 1.228^{***} & 1.190^{***} & 1.565^{**} & 3.472^{***} & 3.368^{***} \\
 & (0.446) & (0.304) & (0.331) & (0.629) & (0.510) & (0.526) \\
 & [0.868] & [0.643] & [0.709] & [2.084] & [1.497] & [1.596] \\
\addlinespace[4pt]
Pre-trend $F$ & 1.19 & 0.69 & 0.67 & 0.71 & 0.29 & 0.28 \\
 & [0.31] & [0.63] & [0.65] & [0.61] & [0.92] & [0.92] \\
\midrule
Pre-treatment mean & \multicolumn{1}{c}{1.26} & \multicolumn{1}{c}{1.26} & \multicolumn{1}{c}{1.26} & \multicolumn{1}{c}{4.69} & \multicolumn{1}{c}{4.69} & \multicolumn{1}{c}{4.69} \\
$N$ (treated) & \multicolumn{1}{c}{543} & \multicolumn{1}{c}{543} & \multicolumn{1}{c}{543} & \multicolumn{1}{c}{543} & \multicolumn{1}{c}{543} & \multicolumn{1}{c}{543} \\
$N$ (control) & \multicolumn{1}{c}{41} & \multicolumn{1}{c}{41} & \multicolumn{1}{c}{41} & \multicolumn{1}{c}{41} & \multicolumn{1}{c}{41} & \multicolumn{1}{c}{41} \\
\midrule
Deal FE & \multicolumn{1}{c}{\checkmark} & \multicolumn{1}{c}{\checkmark} & \multicolumn{1}{c}{\checkmark} & \multicolumn{1}{c}{\checkmark} & \multicolumn{1}{c}{\checkmark} & \multicolumn{1}{c}{\checkmark} \\
Year FE & \multicolumn{1}{c}{\checkmark} &  &  & \multicolumn{1}{c}{\checkmark} &  &  \\
Country $\times$ Year FE &  & \multicolumn{1}{c}{\checkmark} & \multicolumn{1}{c}{\checkmark} &  & \multicolumn{1}{c}{\checkmark} & \multicolumn{1}{c}{\checkmark} \\
Rainfall control &  &  & \multicolumn{1}{c}{\checkmark} &  &  & \multicolumn{1}{c}{\checkmark} \\
\bottomrule
\end{tabular}
\begin{tablenotes}[para,flushleft]
\footnotesize
\item \textit{Notes.} Each column is a separate event study using the \citet{borusyak2024revisiting} imputation estimator. The unit of observation is an LSLA deal, with outcomes aggregated within a 50\,km buffer. The baseline comparison sample consists of implemented deals ($N=543$) and failed deals ($N=41$); the imputation uses all untreated observations, including pre-treatment observations of eventually treated units. The protests outcome counts ACLED ``Protests'' and ``Riots'' events within the buffer. The all-conflict outcome counts all ACLED event types (Battles, Protests, Riots, Violence against civilians, Explosions/Remote violence, and Strategic developments). Bootstrapped standard errors (200 iterations) clustered by deal are reported in parentheses, and \citet{conley1999gmm} spatial HAC standard errors are reported in brackets (Bartlett kernel, 200\,km bandwidth). ATT is the average post-treatment effect ($k \geq 0$). Pre-trend $F$ is the average squared $t$-statistic for $k < 0$; joint $p$-values are reported in brackets. \\
\item $^{*}p<0.10$, $^{**}p<0.05$, $^{***}p<0.01$
\end{tablenotes}
\end{threeparttable}
\end{table}

%% file: tables/A11_spatial_accuracy_later_treated.tex
\begin{table}[tbp]
\centering
\caption{Later-Treated Timing Robustness at $50\,\text{km}$}
\label{tab:spatial_accuracy_later_treated}
\begin{threeparttable}
\setlength{\tabcolsep}{8pt}
\begin{tabular}{lcccc}
\toprule
 & \multicolumn{2}{c}{Protests} & \multicolumn{2}{c}{All conflict events} \\
\cmidrule(lr){2-3} \cmidrule(lr){4-5}
 & \shortstack{(1)\\Failed-deal\\controls} & \shortstack{(2)\\Later-treated\\controls} & \shortstack{(3)\\Failed-deal\\controls} & \shortstack{(4)\\Later-treated\\controls} \\
\midrule
ATT (post-treatment) & $1.228^{***}$ & $1.377^{***}$ & $3.472^{***}$ & $2.182^{***}$ \\
 & (0.304) & (0.295) & (0.510) & (0.382) \\
\addlinespace[4pt]
Pre-trend $F$ & 0.86 & 1.60 & 0.23 & 0.83 \\
 & [0.49] & [0.17] & [0.92] & [0.51] \\
\midrule
Pre-treatment mean & 1.26 & 1.03 & 4.69 & 2.91 \\
Implemented deals & 543 & 482 & 543 & 482 \\
Failed controls & 41 & --- & 41 & --- \\
Cohort stacks & --- & 28 & --- & 28 \\
\midrule
Deal FE & \checkmark & \checkmark & \checkmark & \checkmark \\
Country $\times$ Year FE & \checkmark & \checkmark & \checkmark & \checkmark \\
Rainfall control & \checkmark & \checkmark & \checkmark & \checkmark \\
Treatment onset & Negotiation & Negotiation & Negotiation & Negotiation \\
Control lag window & --- & 5--10 years & --- & 5--10 years \\
\bottomrule
\end{tabular}
\begin{tablenotes}[para,flushleft]
\footnotesize
\item \textit{Notes.} All columns use the \citet{borusyak2024revisiting} imputation estimator on cumulative ACLED outcomes aggregated within a $50\,\text{km}$ buffer, with deal fixed effects, country $\times$ year fixed effects, negotiation-year treatment onset, rainfall controls, and 200 bootstrap iterations clustered at the deal level. Columns (1) and (3) report the preferred implemented-versus-failed Donaldson design from the main text. Columns (2) and (4) hold that estimator and outcome construction fixed but replace the failed-deal counterfactual with a stacked later-treated comparison in which each treated cohort is compared to deals implemented 5 to 10 years later, using only those later cohorts' not-yet-treated observations. The later-treated sample therefore contains only eventually implemented deals; the same deal can serve as a control for earlier cohorts and as treated for its own cohort. ATT is the average post-treatment effect for event times $k \geq 0$. Pre-trend $F$ is the average squared $t$-statistic for pre-treatment leads; joint $p$-values are reported in brackets.
\item $^{*}p<0.10$, $^{**}p<0.05$, $^{***}p<0.01$
\end{tablenotes}
\end{threeparttable}
\end{table}

%% file: tables/A12_early_cohort_exclusion.tex
\begin{table}[tbp]
\centering
\caption{Early-Cohort Exclusion Robustness Check (Drop Negotiation Year $<2003$)}
\label{tab:early_cohort_exclusion_check}
\begin{threeparttable}
\setlength{\tabcolsep}{10pt}
\begin{tabular}{lD{.}{.}{3}D{.}{.}{3}D{.}{.}{3}D{.}{.}{3}D{.}{.}{3}D{.}{.}{3}}
\toprule
\textit{Dependent Variable:} & \multicolumn{3}{c}{Protests} & \multicolumn{3}{c}{All conflict events} \\
\cmidrule(lr){2-4} \cmidrule(lr){5-7}
 & \multicolumn{1}{c}{(1)} & \multicolumn{1}{c}{(2)} & \multicolumn{1}{c}{(3)} & \multicolumn{1}{c}{(4)} & \multicolumn{1}{c}{(5)} & \multicolumn{1}{c}{(6)} \\
\midrule
\multicolumn{7}{l}{\textit{Panel A: Full sample}} \\
ATT (post-treatment) & 1.106^{***} & 1.438^{***} & 1.484^{***} & 2.904^{***} & 2.915^{***} & 3.080^{***} \\
 & (0.207) & (0.192) & (0.200) & (0.334) & (0.351) & (0.372) \\
 & [0.640] & [0.587] & [0.624] & [1.560] & [1.297] & [1.336] \\
\addlinespace[4pt]
Pre-trend $F$ & 1.89 & 1.33 & 1.38 & 0.64 & 0.66 & 0.66 \\
 & [0.09] & [0.25] & [0.23] & [0.67] & [0.65] & [0.65] \\
\midrule
Pre-treatment mean & \multicolumn{1}{c}{0.94} & \multicolumn{1}{c}{0.94} & \multicolumn{1}{c}{0.94} & \multicolumn{1}{c}{3.90} & \multicolumn{1}{c}{3.90} & \multicolumn{1}{c}{3.90} \\
$N$ (treated) & \multicolumn{1}{c}{1,107} & \multicolumn{1}{c}{1,107} & \multicolumn{1}{c}{1,107} & \multicolumn{1}{c}{1,107} & \multicolumn{1}{c}{1,107} & \multicolumn{1}{c}{1,107} \\
$N$ (control) & \multicolumn{1}{c}{284} & \multicolumn{1}{c}{284} & \multicolumn{1}{c}{284} & \multicolumn{1}{c}{284} & \multicolumn{1}{c}{284} & \multicolumn{1}{c}{284} \\
\midrule
\multicolumn{7}{l}{\textit{Panel B: Drop pre-2003}} \\
ATT (post-treatment) & 1.716^{***} & 1.471^{***} & 1.531^{***} & 3.486^{***} & 2.581^{***} & 2.756^{***} \\
 & (0.247) & (0.258) & (0.291) & (0.425) & (0.500) & (0.500) \\
 & [0.752] & [0.685] & [0.724] & [1.830] & [1.484] & [1.526] \\
\addlinespace[4pt]
Pre-trend $F$ & 2.75 & 0.99 & 0.95 & 1.15 & 0.99 & 0.97 \\
 & [0.02] & [0.42] & [0.45] & [0.33] & [0.42] & [0.43] \\
\midrule
Pre-treatment mean & \multicolumn{1}{c}{0.96} & \multicolumn{1}{c}{0.96} & \multicolumn{1}{c}{0.96} & \multicolumn{1}{c}{3.97} & \multicolumn{1}{c}{3.97} & \multicolumn{1}{c}{3.97} \\
$N$ (treated) & \multicolumn{1}{c}{766} & \multicolumn{1}{c}{766} & \multicolumn{1}{c}{766} & \multicolumn{1}{c}{766} & \multicolumn{1}{c}{766} & \multicolumn{1}{c}{766} \\
$N$ (control) & \multicolumn{1}{c}{215} & \multicolumn{1}{c}{215} & \multicolumn{1}{c}{215} & \multicolumn{1}{c}{215} & \multicolumn{1}{c}{215} & \multicolumn{1}{c}{215} \\
\midrule
Deal FE & \multicolumn{1}{c}{\checkmark} & \multicolumn{1}{c}{\checkmark} & \multicolumn{1}{c}{\checkmark} & \multicolumn{1}{c}{\checkmark} & \multicolumn{1}{c}{\checkmark} & \multicolumn{1}{c}{\checkmark} \\
Year FE & \multicolumn{1}{c}{\checkmark} &   &   & \multicolumn{1}{c}{\checkmark} &   &   \\
Country $\times$ Year FE &   & \multicolumn{1}{c}{\checkmark} & \multicolumn{1}{c}{\checkmark} &   & \multicolumn{1}{c}{\checkmark} & \multicolumn{1}{c}{\checkmark} \\
Rainfall control &   &   & \multicolumn{1}{c}{\checkmark} &   &   & \multicolumn{1}{c}{\checkmark} \\
\bottomrule
\end{tabular}
\begin{tablenotes}[para,flushleft]
\footnotesize
\item \textit{Notes.} Early-cohort exclusion robustness check. Panel A reproduces the baseline analytical sample. Panel B drops all implemented and failed deals whose negotiation year is earlier than 2003, restricting the sample to cohorts whose pre-treatment windows fall entirely within the period of mature ACLED coverage. Both panels use the \citet{borusyak2024revisiting} imputation estimator on a $50\,\text{km}$ buffer with negotiation-year treatment onset; columns vary the time fixed effects and rainfall control as in the main baseline. Bootstrapped standard errors (200 iterations) clustered by deal are reported in parentheses; \citet{conley1999gmm} spatial HAC standard errors are reported in brackets (Bartlett kernel, 200\,km bandwidth). Pre-trend $F$ is the average squared $t$-statistic across pre-treatment leads; joint $p$-values are reported in brackets. \\
\item $^{*}p<0.10$, $^{**}p<0.05$, $^{***}p<0.01$
\end{tablenotes}
\end{threeparttable}
\end{table}

%% file: tables/A13_buffer_overlap.tex
\begin{table}[tbp]
\centering
\caption{Buffer Overlap Among Treated Deals}
\label{tab:buffer_overlap}
\begin{threeparttable}
\setlength{\tabcolsep}{8pt}
\begin{tabular}{lcccc}
\toprule
Buffer & Deals & Deals w/ overlap & Share & Mean neighbors \\
\midrule
  25km & 1,107 & 879 & 79.4\% & 6.0 \\
  50km & 1,107 & 1,019 & 92.1\% & 10.4 \\
  75km & 1,107 & 1,061 & 95.8\% & 17.4 \\
\bottomrule
\end{tabular}
\begin{tablenotes}[para,flushleft]
\footnotesize
\item \textit{Notes.} Two deals' buffers overlap when the distance between their coordinates is less than twice the buffer radius. ``Deals w/ overlap'' counts treated deals that have at least one other treated deal within this threshold. ``Mean neighbors'' is the average number of overlapping partners among deals with any overlap. Sample: implemented deals in the baseline panel.
\end{tablenotes}
\end{threeparttable}
\end{table}

%% file: tables/A14_drop_clustered.tex
\begin{table}[tbp]
\centering
\caption{Dropping High-Overlap Treated Deals}
\label{tab:drop_clustered}
\begin{threeparttable}
\setlength{\tabcolsep}{6pt}
\begin{tabular}{lcccc}
\toprule
 & (1) & (2) & (3) & (4) \\
 & Baseline & Drop top decile & Drop $>10$ & Drop $>5$ \\
\midrule
 ATT & $1.438^{***}$ & $1.404^{***}$ & $1.308^{***}$ & $1.684^{***}$ \\
 & (0.192) & (0.202) & (0.260) & (0.362) \\
\addlinespace
\midrule
 Overlap threshold & --- & $>20$ & $>10$ & $>5$ \\
 $N$ (treated) & 1,107 & 1,016 & 693 & 447 \\
 $N$ (control) & 284 & 284 & 284 & 284 \\
\midrule
 Deal FE & \checkmark & \checkmark & \checkmark & \checkmark \\
 Country $\times$ Year FE & \checkmark & \checkmark & \checkmark & \checkmark \\
\bottomrule
\end{tabular}
\begin{tablenotes}[para,flushleft]
\footnotesize
\item \textit{Notes.} Pooled post-treatment ATT from the \citet{borusyak2024revisiting} imputation estimator. Outcome: annual count of civic protests and riots within a 50\,km buffer. For each treated deal we count the number of other treated deals whose centroid lies within $100$\,km (the distance at which two 50\,km buffers overlap). Column (1) uses the full sample. Column (2) drops treated deals in the top decile of neighbor count (threshold $>20$). Columns (3) and (4) drop deals with more than 10 and 5 neighbors, respectively. Controls are unaffected. Cluster bootstrap SE (200 iterations, clustered at deal) in parentheses. \\
\item $^{*}p<0.10$, $^{**}p<0.05$, $^{***}p<0.01$
\end{tablenotes}
\end{threeparttable}
\end{table}

%% file: tables/A15_armed_conflict_outcomes.tex
\begin{table}[tbp]
\centering
\caption{Parallel Armed-Conflict Outcomes Around LSLA Negotiations}
\label{tab:armed_conflict_outcomes}
\begin{threeparttable}
\small
\begin{tabular}{lccccc}
\toprule
Outcome & ATT & SE & 95\% CI & Pre-trend $F$ & Pre-mean \\
\midrule
Protests & 1.060*** & 0.142 & [0.782, 1.337] & 1.42 & 0.60 \\
Armed conflict & 1.256*** & 0.216 & [0.833, 1.680] & 1.24 & 2.23 \\
Battles & 0.624*** & 0.095 & [0.437, 0.811] & 0.83 & 1.11 \\
Explosions / remote violence & 0.120*** & 0.017 & [0.086, 0.153] & 1.04 & 0.08 \\
Violence against civilians & 0.513*** & 0.132 & [0.255, 0.771] & 1.34 & 1.03 \\
Strategic developments & 0.220*** & 0.058 & [0.107, 0.334] & 2.14 & 0.73 \\
Fatalities & 9.577*** & 2.722 & [4.241, 14.912] & 1.03 & 12.03 \\
\bottomrule
\end{tabular}
\begin{tablenotes}[para,flushleft]
\footnotesize
\item \textit{Notes.} Armed conflict is defined as the sum of ACLED Battles, Explosions/Remote violence, and Violence against civilians. Strategic developments and fatalities are reported separately as diagnostics. Each row reports the aggregate post-treatment ATT from the \citet{borusyak2024revisiting} imputation estimator with unit and country $\times$ year fixed effects. Standard errors are cluster bootstrap estimates.
\item $^{*}p<0.10$, $^{**}p<0.05$, $^{***}p<0.01$
\end{tablenotes}
\end{threeparttable}
\end{table}

%% file: tables/A16_heterogeneity_full.tex
\begin{table}[tbp]
\centering
\caption{Later-Treated Land-Ownership Robustness under Broader Definitions}
\label{tab:heterogeneity_full}
\begin{threeparttable}
\setlength{\tabcolsep}{12pt}
\begin{tabular}{lD{.}{.}{3}D{.}{.}{3}D{.}{.}{3}D{.}{.}{3}D{.}{.}{3}D{.}{.}{3}}
\toprule
\textit{Dependent Variable:} & \multicolumn{3}{c}{Protests} & \multicolumn{3}{c}{All conflict events} \\
\cmidrule(lr){2-4} \cmidrule(lr){5-7}
 & \multicolumn{1}{c}{(1)} & \multicolumn{1}{c}{(2)} & \multicolumn{1}{c}{(3)} & \multicolumn{1}{c}{(4)} & \multicolumn{1}{c}{(5)} & \multicolumn{1}{c}{(6)} \\
\midrule
\multicolumn{7}{l}{\textit{Panel A: Any community/indigenous mention}} \\
ATT (post-treatment) & 2.624^{***} & 3.621^{***} & 3.665^{***} & 4.397^{***} & 5.152^{***} & 4.566^{***} \\
 & (0.469) & (0.437) & (0.455) & (0.711) & (0.633) & (0.713) \\
 & [1.099] & [1.185] & [1.201] & [1.701] & [1.812] & [1.789] \\
\addlinespace[4pt]
Pre-trend $F$ & 1.68 & 0.35 & 0.37 & 0.69 & 1.01 & 0.86 \\
 & [0.13] & [0.88] & [0.87] & [0.63] & [0.41] & [0.50] \\
\midrule
Pre-treatment mean & \multicolumn{1}{c}{0.48} & \multicolumn{1}{c}{0.48} & \multicolumn{1}{c}{0.48} & \multicolumn{1}{c}{4.88} & \multicolumn{1}{c}{4.88} & \multicolumn{1}{c}{4.88} \\
$N$ (implemented) & \multicolumn{1}{c}{131} & \multicolumn{1}{c}{131} & \multicolumn{1}{c}{131} & \multicolumn{1}{c}{131} & \multicolumn{1}{c}{131} & \multicolumn{1}{c}{131} \\
Cohort stacks & \multicolumn{1}{c}{19} & \multicolumn{1}{c}{19} & \multicolumn{1}{c}{19} & \multicolumn{1}{c}{19} & \multicolumn{1}{c}{19} & \multicolumn{1}{c}{19} \\
\midrule
\multicolumn{7}{l}{\textit{Panel B: Any state mention}} \\
ATT (post-treatment) & 1.664^{***} & 1.861^{***} & 1.933^{***} & 1.208^{**} & 1.351^{***} & 0.674 \\
 & (0.208) & (0.205) & (0.249) & (0.608) & (0.500) & (0.565) \\
 & [0.863] & [0.894] & [0.939] & [1.572] & [1.460] & [1.485] \\
\addlinespace[4pt]
Pre-trend $F$ & 1.99 & 1.73 & 1.46 & 1.57 & 0.91 & 0.68 \\
 & [0.08] & [0.12] & [0.20] & [0.16] & [0.47] & [0.64] \\
\midrule
Pre-treatment mean & \multicolumn{1}{c}{0.36} & \multicolumn{1}{c}{0.36} & \multicolumn{1}{c}{0.36} & \multicolumn{1}{c}{3.09} & \multicolumn{1}{c}{3.09} & \multicolumn{1}{c}{3.09} \\
$N$ (implemented) & \multicolumn{1}{c}{147} & \multicolumn{1}{c}{147} & \multicolumn{1}{c}{147} & \multicolumn{1}{c}{147} & \multicolumn{1}{c}{147} & \multicolumn{1}{c}{147} \\
Cohort stacks & \multicolumn{1}{c}{21} & \multicolumn{1}{c}{21} & \multicolumn{1}{c}{21} & \multicolumn{1}{c}{21} & \multicolumn{1}{c}{21} & \multicolumn{1}{c}{21} \\
\midrule
\multicolumn{7}{l}{\textit{Panel C: Community or smallholders (any mention)}} \\
ATT (post-treatment) & 2.544^{***} & 3.535^{***} & 3.781^{***} & 4.675^{***} & 5.145^{***} & 4.932^{***} \\
 & (0.416) & (0.403) & (0.435) & (0.595) & (0.556) & (0.645) \\
 & [1.095] & [1.136] & [1.171] & [1.729] & [1.777] & [1.786] \\
\addlinespace[4pt]
Pre-trend $F$ & 2.67 & 0.89 & 0.96 & 0.41 & 0.78 & 0.73 \\
 & [0.02] & [0.49] & [0.44] & [0.84] & [0.57] & [0.60] \\
\midrule
Pre-treatment mean & \multicolumn{1}{c}{0.69} & \multicolumn{1}{c}{0.69} & \multicolumn{1}{c}{0.69} & \multicolumn{1}{c}{4.94} & \multicolumn{1}{c}{4.94} & \multicolumn{1}{c}{4.94} \\
$N$ (implemented) & \multicolumn{1}{c}{149} & \multicolumn{1}{c}{149} & \multicolumn{1}{c}{149} & \multicolumn{1}{c}{149} & \multicolumn{1}{c}{149} & \multicolumn{1}{c}{149} \\
Cohort stacks & \multicolumn{1}{c}{21} & \multicolumn{1}{c}{21} & \multicolumn{1}{c}{21} & \multicolumn{1}{c}{21} & \multicolumn{1}{c}{21} & \multicolumn{1}{c}{21} \\
\midrule
Deal FE & \multicolumn{1}{c}{\checkmark} & \multicolumn{1}{c}{\checkmark} & \multicolumn{1}{c}{\checkmark} & \multicolumn{1}{c}{\checkmark} & \multicolumn{1}{c}{\checkmark} & \multicolumn{1}{c}{\checkmark} \\
Year FE & \multicolumn{1}{c}{\checkmark} &   &   & \multicolumn{1}{c}{\checkmark} &   &   \\
Country $\times$ Year FE &   & \multicolumn{1}{c}{\checkmark} & \multicolumn{1}{c}{\checkmark} &   & \multicolumn{1}{c}{\checkmark} & \multicolumn{1}{c}{\checkmark} \\
Rainfall control &   &   & \multicolumn{1}{c}{\checkmark} &   &   & \multicolumn{1}{c}{\checkmark} \\
\bottomrule
\end{tabular}
\begin{tablenotes}[para,flushleft]
\footnotesize
\item \textit{Notes.} Robustness: broader, non-mutually-exclusive ownership definitions. Panel A includes any deal with a community or indigenous mention in the prior-owner field; Panel B includes any deal with a state mention; Panel C includes any deal with a community or smallholder mention. Mixed labels such as State$|$Community or Private (smallholders)$|$Community contribute to the relevant categories, so deals can appear in multiple panels. Estimation, controls, and standard errors are implemented analogous to Table~\ref{tab:ownership_later_treated_main}. \\
\item $^{*}p<0.10$, $^{**}p<0.05$, $^{***}p<0.01$
\end{tablenotes}
\end{threeparttable}
\end{table}